\documentclass[aps,nofootinbib,longbibliography,superscriptaddress,
tightenlines,notitlepage,prd]{revtex4-1}

\usepackage{amsmath,amsthm,latexsym,amssymb,amsfonts}
\usepackage{graphicx,color,natbib}
\usepackage[export]{adjustbox}
\usepackage[colorlinks=true]{hyperref}

\date{\today}
\numberwithin{equation}{section}

\begin{document}
	
\title{Numerical evaluation of spin foam amplitudes beyond simplices}

\author{Courtney Allen}
\email{callen15@uoguelph.ca}
\affiliation{Department of Mathematics \& Statistics,
	University of Guelph,\\	50 Stone Road East,
	Guelph, Ontario, Canada,
	N1G 2W1}
\author{Florian Girelli}
\email{florian.girelli@uwaterloo.ca}
\affiliation{Department of Applied Mathematics, University of Waterloo,\\ 200 University Avenue West, Waterloo, Ontario, Canada, N2L 3G1}
\author{Sebastian Steinhaus}
\email{sebastian.steinhaus@uni-jena.de}
\affiliation{Theoretisch-Physikalisches Institut, Friedrich-Schiller-Universit\"at Jena,\\ Max-Wien-Platz 1, 07743 Jena, Germany}

\begin{abstract}
	We present the first numerical calculation of the 4D Euclidean spin foam vertex amplitude for vertices with hypercubic combinatorics. Concretely, we compute the amplitude for coherent boundary data peaked on cuboid and frustum shapes. We present the numerical algorithms to explicitly compute the vertex amplitude and compare the results in different cases to the semi-classical approximation of the amplitude. Overall we find good qualitative agreement of the amplitudes and evidence of convergence of the asymptotic formula to the full amplitude already at fairly small spins, yet also differences in the frequency of oscillations and a phase shift absent in the 4-simplex case. However, due to rapidly growing numerical costs, we cannot reach sufficiently high spins to prove agreement of both amplitudes.  Lastly, this setup allows us to explore non-uniform vertex amplitudes, where some representations are small while others are large; we find indications that scenarios might exist in which the semi-classical amplitude is a valid approximation even if some spins remain small. This suggests that the transition of the quantum to the semi-classical regime (for a single vertex amplitude) is intricate.
\end{abstract}

\maketitle


\section{Introduction}

Spin foam quantum gravity \cite{carlobook,Perez:2012wv} aims at defining a non-perturbative, background independent quantization of general relativity as a path integral over geometries closely related to canonical loop quantum gravity \cite{thomasbook}. 
The currently most advanced and most frequently used spin foam model in 4D is the so-called Engle-Pereira-Rovelli-Livine / Freidel-Krasnov model\cite{Engle:2007uq,Engle:2007qf,Engle:2007wy,Freidel:2007py}, EPRL-FK for short. One of its advantages is that it makes direct contact to the kinematical Hilbert space of loop quantum gravity, via projected spin network states \cite{Livine:2002ak}, and defines an evolution of spin network states. To do so, the boundary graph is extended one dimension higher to a 2-complex, which is colored by the same group theoretic data as the spin network, i.e. group representations and intertwiners. We implement a dynamics by locally assigning an amplitude to each spin foam state and then summing over the bulk geometries encoded in the bulk data. Originally, the EPRL-FK model is defined for 2-complexes dual to triangulations, but Kaminski, Kiesilowski and Lewandowski (KKL) extended the theory to arbitrary 2-complexes in \cite{Kaminski:2009fm} to allow for arbitrary boundary graphs.

A key role in any spin foam calculation is played by the vertex amplitude assigned to (vertices dual to) the 4-dimensional building blocks of the theory. This amplitude is thoroughly studied and well-understood in the so-called semi-classical regime: if all group representations assigned to a vertex are ``large'', the vertex amplitude is well approximated by a stationary phase approximation. The critical points dominating this approximation correspond to Regge geometries, triangulations fully prescribed by the lengths of their edges, and weighted by the exponential of the Regge action \cite{Regge:1961px}, a discretization of general relativity. With the exception of so-called vector geometries\footnote{The coherent 4-simplex amplitude is labeled by ten spins and 20 3d normal vectors. If all tetrahedra close, we have so-called twisted geometries \cite{Freidel:2010aq} parametrised by (gauge-invariant) 20 variables. Additionally, if the tetrahedra can be individually rotated such that their normals on shared triangles are pairwise anti-parallel, we have a vector geometry; these span a 15-dimensional subspace. Note that the triangles seen from different tetrahedra need not have the same shape. Enforcing shape matching leads to the 10-dimensioanl subspace of Regge geometries, from which flat Euclidean 4-simplices are reconstructed. See \cite{Dona:2017dvf} for a thorough explanation of these geometries in the Euclidean setting.} \cite{Barrett:2009as,Dona:2017dvf}, all other geometries are exponentially suppressed in this limit. These results are universally derived in spin foams, e.g. for different Euclidean models \cite{Barrett:1998gs,Conrady:2008mk,Barrett2009a}, Lorentzian signature (with space-like and time-like tetrahedra and triangles) \cite{Barrett2011,Kaminski2018a,Liu2019,Simao:2021qno} and for vertices more general than 4-simplices \cite{Bahr:2015gxa,Bahr:2017eyi,Dona:2017dvf,Dona:2020yao,Assanioussi:2020fml}.

Unfortunately, for small representations, which we will call the quantum regime, the asymptotic expansion breaks down without other analytical methods to replace it. In recent years there has been significant progress to close this gap using numerical methods \cite{Dona:2017dvf,Dona:2019dkf,Dona:2019jab,Gozzini:2021kbt}: these articles describe in detail the development and optimization of a numerical package \verb|sl2cfoam|\footnote{\url{https://github.com/qg-cpt-marseille/sl2cfoam}} (and \verb|slc2foam-next|\footnote{\url{https://github.com/qg-cpt-marseille/sl2cfoam-next}}), which computes the 4-simplex vertex amplitude of the Lorentzian EPRL model making use of an important identity of $\text{SL}(2,\mathbb{C})$ group elements derived in \cite{Speziale:2016axj}. These algorithms furthermore lead to first studies of 2-complexes with several 4-simplices \cite{Dona:2020tvv}, e.g. to explore the so-called ``flatness problem'' in spin foams \cite{Bonzom:2009hw,Hellmann:2013gva,Oliveira:2017osu,Engle:2020ffj,Asante:2020iwm,Han:2021kll}, yet they remain challenging due to rapidly growing numerical costs as the representation labels are increased. Additionally, these promising results are complemented by developments that will help to unlock larger 2-complexes: effective spin foam models \cite{Asante:2020qpa,Asante:2020iwm,Asante:2021zzh} bypass this numerical challenge by directly starting from the asymptotic formula to investigate under which conditions reasonable semi-classical physics emerge, while Lefshetz thimbles enable the use of Monte Carlo methods to compute expectation values of observables on larger 2-complexes \cite{Han:2020npv}.

In this article we present, to our knowledge, the first calculation of Euclidean 4D vertex amplitudes for a 2-complex more general than a triangulation, here with combinatorics of a hybercubic lattice. We numerically compute the amplitude for specific coherent boundary data corresponding to cuboids and frusta; the corresponding asymptotic expansions were computed in \cite{Bahr:2015gxa} and \cite{Bahr:2017eyi} respectively\footnote{See \cite{Bahr:2018ewi} for an inclusion of a non-vanishing cosmological constant into the frustum model.}.
In both cases, the vertex amplitude is defined as the contraction of eight six-valent intertwiners. In the cuboid case, the intertwiners are peaked on the shape of cuboids, resulting in a vertex dual to a (not necessarily shape-matching \cite{Bahr:2015gxa,Dona:2017dvf}) hypercuboid. For frusta, we have different types of intertwiners: two cubes of different size connected by frusta\footnote{A frustum can be imagined as a pyramid with a square base, whose tip has been cut off parallel to its base.}. This model can thus describe expanding or contracting cubulations.
The cuboid or frustum data were chosen specifically to define restricted models: instead of summing over all possible intertwiners, the models were restricted by hand in order to explore a subset of the full gravitational path integral. The restrictions, together with using the semi-classical approximation and the regular combinatorics of the 2-complex, opened the door to numerically derive the first renormalization group flow of 4D spin foam models \cite{Bahr:2016hwc,Bahr:2017klw,Bahr:2018gwf}, which revealed indications for UV-attractive fixed point at which diffeomorphism symmetry could be restored. Furthermore, the spectral dimension of the cuboid model was investigated in \cite{Steinhaus:2018aav} for periodic spin foams, which revealed a mechanism for how the superposition of geometries of different size can lead to a reduction in this effective dimension measure.

This article is organized as follows: in section \ref{sec:restricted_sf} we give a brief introduction to spin foam models, more precisely the Euclidean EPRL-FK model. We discuss the KKL-extension to more general 2-complexes and introduce the coherent cuboid and frustum states for the intertwiners. We close the section by recapping the derivation of the asymptotic formula of the vertex amplitude. Section \ref{sec:numerical} details the numerical computation of the vertex amplitude, from the definition of the intertwiners in cuboid and frustum case and their contraction giving the vertex amplitude. Furthermore, we describe the optimization of the code and its parallelization. In section \ref{sec:results} we present the results for various examples in both cuboid and frustum cases and compare the full amplitude to its asymptotic expansion. We conclude with a discussion and outlook in section \ref{sec:discussion}. As supplementary materials, the code\footnote{\url{https://github.com/CourtA96/SpinfoamAmplitudesOfCuboidsAndFrustra}} and data from simulations\footnote{\url{https://doi.org/10.5281/zenodo.6006163}} are openly available.

\section{Restricted spin foam models} \label{sec:restricted_sf}

The goal of this article is to numerically compute the vertex amplitude of restricted spin foam models defined in \cite{Bahr:2015gxa} and \cite{Bahr:2017eyi} and compare it to its semi-classical approximation. These models were defined for the Euclidean EPRL-FK model for Barbero-Immirzi parameter $\gamma < 1$ on a hypercubic 2-complex, following the Kaminski-Kisielowski-Lewandowski (KKL) extension \cite{Kaminski:2009fm}. In the following, we give a brief introduction to spin foam models, present the restricted models and set the scope of this article.

The EPRL-FK model \cite{Engle:2007qf,Engle:2007uq,Engle:2007wy,Freidel:2007py} is originally defined on 2-complexes dual to 4d triangulations. Such a 2-complex consists of vertices $v$, edges $e$ and faces $f$, which are dual to 4-simplices, tetrahedra and triangles respectively in the dual triangulation. 
Part of the definition is equipping the 2-complex with fiducial orientations on its edges and faces, which do not influence the results.

A spin foam state is given by a coloring of the 2-complex with group theoretic data (here for $\text{SU}(2)$): To each face $f$ of the 2-complex one assigns an irreducible representation of the symmetry group, e.g. $\text{SU}(2)$ representations labeled by a spin $j_f \in \frac{\mathbb{N}}{2}$. Each edge $e$ carries an intertwiner $\iota_e$, an invariant tensor under the action of the group in the product space of representation vector spaces, e.g. $V_{j_1} \otimes \dots \otimes V_{j_4}$ for a 4-valent $\text{SU}(2)$ intertwiner dual to a tetrahedron. The spins $j_1$ to $j_4$ are associated to the faces, which share the edge $e$. Geometrically, we interpret these data as follows: in the orthonormal spin network basis, a 4-valent intertwiner is labeled by five $\text{SU}(2)$ spins, four assigned to its faces and one recoupling label. These correspond to the areas of the four triangles of the tetrahedron as well as the area of a parallelogram inside the tetrahedron determined by the recoupling scheme. In contrast to classical tetrahedra, which are uniquely specified (up to rotations, translations, etc.) by six edge lengths, these polyhedra are quantum \cite{Baez:1999tk}. Using coherent states \cite{Perelomov_1986} we can define intertwiners that are sharply peaked on classical polyhedra, which play a central role in the definition of the restricted spin foam models and the computation of their asymptotic amplitudes.

The assignment of amplitudes to spin foam states defines a spin foam model; this assignment is local, i.e. we assign an amplitude to each face, edge and vertex of the 2-complex, which then only depends on the group theoretic data adjacent to that object. For example, the face amplitude solely depends on the representation associated to that face. Given all these ingredients, the spin foam partition function is then given as a sum over all spin foam states, i.e. all irreducible representations and intertwiners, weighted by the spin foam amplitudes:
\begin{equation}
	Z = \sum_{j_f,\iota_e} \prod_f \mathcal{A}_f \prod_e \mathcal{A}_e \prod_v \mathcal{A}_v \quad .
\end{equation}
The symbols $\mathcal{A}_f$, $\mathcal{A}_e$ and $\mathcal{A}_v$ denote the face, edge and vertex amplitudes respectively, where our main focus is on the vertex amplitude in this article.

The Euclidean EPRL-FK model \cite{Engle:2007wy,Engle:2007uq,Freidel:2007py} is derived by imposing simplicity constraints onto $\text{Spin}(4) \simeq \text{SU}(2) \times \text{SU}(2)$ BF theory to break the topological symmetry of this theory. In this model, these constraints restrict the $\text{Spin}(4)$ representations $(j^+,j^-)$, where $j^\pm$ label $\text{SU}(2)$ spins. $j^\pm$ must satisfy the condition
\begin{equation}
	j^{\pm} = \frac{1}{2} (1 \pm \gamma) j \quad ,
\end{equation}
where we choose the Barbero-Immirzi parameter $\gamma < 1$ and $j \in \frac{1}{2} \mathbb{N}$ is an $\text{SU}(2)$ representation. Since also $j^\pm \in \frac{1}{2} \mathbb{N}$ $\gamma$ is restricted to a rational number, which is considered a peculiarity of the Euclidean model. Note that alternative impositions in the Euclidean theory avoiding this condition exits \cite{Finocchiaro:2018hks}. Moreover, this condition is absent in the Lorentzian theory \cite{Engle:2007wy}.

For the intertwiners one defines a boosting map $\Phi$, which maps an $\text{SU}(2)$ intertwiner to a $\text{SU}(2) \times \text{SU}(2)$ intertwiner. This map consists of two parts: firstly, starting from an $\text{SU}(2)$ intertwiner, we isometrically embed each vector space $V_j$ into the unique component appearing in the Clebsch-Gordan decomposition of $V_{j^+,j^-} \simeq V_{j^+} \otimes V_{j^-}$ by the map $\beta^\gamma_j$. Since the resulting tensor is not invariant under the action of $\text{SU}(2) \times \text{SU}(2)$, one furthermore acts with the Haar projector $\mathcal{P}$ on this tensor. In conclusion, the map $\Phi$ for a 4-valent intertwiner reads:
\begin{equation}
	\Phi : \text{Inv}_{\text{SU}(2)}(V_{j_1} \otimes \dots \otimes V_{j_4}) \rightarrow \text{Inv}_{\text{SU}(2) \times \text{SU}(2)}(V_{j^+_1,j^-_1}  \otimes \dots \otimes V_{j^+_4,j^-_4}) \quad ,
\end{equation}
\begin{equation}
	\Phi := \mathcal{P} (\beta^\gamma_{j_1} \otimes \dots \otimes \beta^\gamma_{j_4}) \quad .
\end{equation}
With these ingredients, the vertex amplitude is defined as the contraction of intertwiners according to the combinatorics of the 2-complex, often expressed as the vertex trace:
\begin{equation}
	\mathcal{A}_v := \text{Tr}_{e \supset v} (\Phi(\iota_e)) \quad .
\end{equation}
The numerical computation of the Euclidean vertex amplitude and the comparison to its semi-classical approximation are the objectives of this article. The numerical evaluation of the EPRL vertex amplitude defined on triangulations has received a lot of attention over the last few years, in particular for the Lorentzian theory, and has seen significant progress \cite{Dona:2019dkf,Gozzini:2021kbt}. In this article, we extend this discussion to more general 2-complexes, namely those with the combinatorics of a hypercubic lattice, where we compute the vertex amplitude for specific types of intertwiners. These are so-called cuboid and frustum intertwiners, which we will discuss next.

\subsection{Cuboid and frustum spin foams}

To define intertwiners that are sharply peaked on classical discrete geometries we use coherent states, more precisely Perelomov coherent states \cite{Perelomov_1986} for $\text{SU}(2)$, which are then boosted $\text{SU}(2) \times \text{SU}(2)$ ones. These states $|j,\vec{n}\rangle$ are maximum weight states and are labeled by a normalized vector $\vec{n} \in S^2$. They satisfy the following equations
\begin{align}
	\vec{n} \cdot \vec{J} |j,\vec{n} \rangle & = j |j,\vec{n} \rangle \quad , \nonumber \\
	\langle j, \vec{n} | \vec{J} |j,\vec{n} \rangle & = \vec{n} \quad ,
\end{align}
where $\vec{J}$ denotes the vector of generators of $\text{SU}(2)$. Starting from the state $|j,j\rangle =: |j,\vec{e}_z\rangle$ (or alternatively $|j,-j\rangle$), we obtain $|j,\vec{n}\rangle$ by acting with a $\text{SU}(2)$ group element on it, which corresponds to the rotation of $\vec{e}_z$ to $\vec{n}$.
\begin{equation}
	|j,\vec{n}\rangle := D^j(g) |j,j\rangle \quad ,
\end{equation}
where $D^j$ denotes the Wigner-$D$ matrix of representation $j$, and $g \in \text{SU}(2)$ encodes the before-mentioned rotation. This notation is often abbreviated in the literature as $g \triangleright |j,j\rangle$, where $\triangleright$ denotes the action of the group. Note that these coherent states are only defined up to a phase.

A coherent intertwiner is defined as the group averaged tensor product of several such $\text{SU}(2)$ coherent states $\{|j_i,\vec{n}_i\}$, with as many coherent states as the valency of the intertwiner. For it to correspond to a classical polyhedron its normals and representations must satisfy the closure condition
\begin{equation} \label{eq:closure}
	\sum_i j_i \vec{n}_i = 0 \quad ,
\end{equation}
which guarantees by Minkowski's theorem \cite{Bianchi:2010gc} that a unique, Euclidean convex polyhedron with the same faces and outward pointing normals exists. A coherent intertwiner, here for a tetrahedron, is then defined as follows:
\begin{equation}
	\iota_{j_i}^{\vec{n}_i} := \int_{\text{SU}(2)} dg \; \bigotimes_{i=1}^4 g \triangleright |j_i,\vec{n}_i \rangle \quad .
\end{equation}

The restricted spin foam models introduced in \cite{Bahr:2015gxa,Bahr:2017eyi} are defined on 2-complexes with regular hypercubic combinatorics: each spin foam vertex is eight-valent, each intertwiner is six-valent and a vertex is shared by 24 faces. To control this added complexity compared to a triangulation case, the key idea is to restrict the sum over intertwiners in the model to specific coherent intertwiners, which furthermore restrict the representations assigned to the faces. In this article we focus on the cuboid and frustum intertwiners.

\subsubsection{Cuboid intertwiners}

\begin{figure}
	\includegraphics[width=0.75 \textwidth]{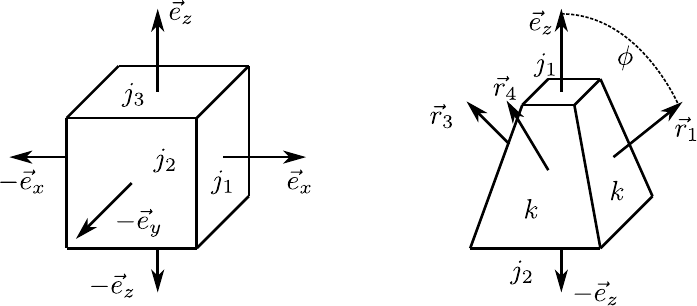}
	\caption{ \label{fig:cuboid_frustum}
		\textit{Left side:} a cuboid characterized by three spins $j_i$ and its outward pointing normals. Spins on opposite faces agree and their normals are anti-parallel. Normals on neighboring faces are orthogonal, here described by Cartesian basis vectors $\vec{e}_i$. \textit{Right side:} A frustum, also characterized by three spins $j_1$, $j_2$ and $k$. $j_1$ and $j_2$ give the area of top and bottom respectively, which each come with anti-parallel normals, here $\pm \vec{e}_z$. The four remaining side faces carry spin $k$; their normals $\vec{r}_l$ enclose the angle $\phi$ with the top normal, which is determined by the spins $j_1$, $j_2$ and $k$. From $\vec{r}_1$, we obtain the remaining normal vectors by rotations of multiples of $\frac{\pi}{2}$ around the $z$-direction.
	}
\end{figure}

\begin{figure}
	\includegraphics[width=0.5\textwidth]{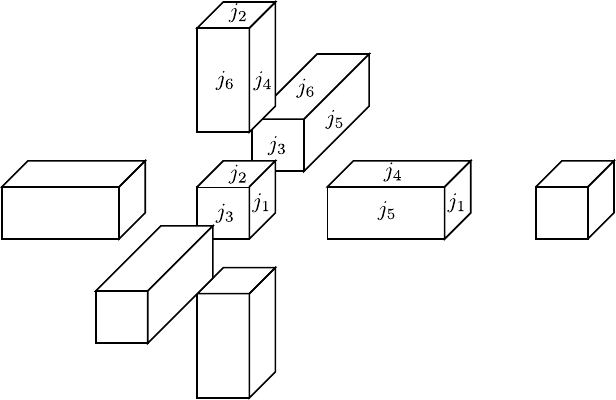}
	\caption{\label{fig:cuboid_bdr}
		The boundary of a 4D cuboid, consisting of eight 3D cuboids. The cuboids are glued along faces that have the same area. Due to the symmetry of cuboids, i.e. opposite faces have the same area, one configuration is determined by six spins / face areas, where we have four types of cuboids.
		}
\end{figure}

Cuboids are straightforward to imagine in terms of face areas and normals, see also fig. \ref{fig:cuboid_frustum}: opposite faces have the same area and anti-parallel normals. Each face is connected to four neighbouring faces, where its normal vector is orthogonal to all normal vectors of adjacent faces. In short, we define this intertwiner by assigning three spins to its three pairs of faces, and choose their normal vectors as the Cartesian basis vectors $\vec{e}_i = \vec{e}_x, \vec{e}_y, \vec{e}_z$.
\begin{equation}
	\iota_{j_1,j_2,j_3} := \int_{\text{SU}(2)} dg \, \bigotimes_{i=1}^3 g \triangleright |j_i,\vec{e}_i\rangle \bigotimes_{i=1}^3 g \triangleright |j_i,-\vec{e}_i\rangle \quad .
\end{equation}
A hypercuboid built from eight such cuboid intertwiners then depends on six representation due to the symmetry of each cuboid intertwiner, see fig. \ref{fig:cuboid_bdr}. Note that while we can compute the three edge lengths of each cuboid from its face areas (if all $j_i > 0$), the edge lengths derived from different cuboids may not agree. These configurations are called angle-matched \cite{Langvik:2016hxn,Dona:2017dvf,Dona:2020yao}, since the area and the angles of the shared face agree, but the face's shape does not. This explains the apparent discrepancy between a classical hypercuboid described by four edge lengths and the vertex amplitude depending on six spins / areas. In \cite{Bahr:2015gxa} it is discussed that these additional degrees of freedom allow for torsionful configurations\footnote{The edge lengths of each 3d cuboid are uniquely determined by its areas, yet the edge lengths of neighbouring cuboids may not agree, such that their face shapes do not match. Thus it is possible to go around a minimal rectangle in the boundary, which does not close due to the mismatch of edge lengths in neighbouring cuboids, matching the definition of torsion in classical continuum gravity.}, and their appearance can be related to the non-implementation of the volume simplicity constraint for the EPRL model defined on 2-complexes more general than triangulations \cite{Bahr:2017ajs,Assanioussi:2020fml}. Shape-matching configurations exist, for which the representation must satisfy the condition (see again fig. \ref{fig:cuboid_bdr}):
\begin{equation}
	j_1 j_6 = j_2 j_5 = j_3 j_4 \quad .
\end{equation}

\subsubsection{Frustum intertwiners}

A frustum can be imagined as a pyramid with square base, whose tip has been cut off (parallel to its base). Moreover, it is a generalization of cuboids, see again fig. \ref{fig:cuboid_frustum}. Consider a cuboid in which two opposite faces carry spin $j$, without loss of generality top and bottom, and all other faces have spin $k$. If we shrink the top spin to $j_1$ we find a frustum as the 3D analogue of a trapezoid, where we require that top and bottom face are squares. Given the face areas of a frustum, i.e. $j_1$, $j_2$ and $k$, its shape is completely specified and can be translated into three edge lengths (one for each square and one for sides of the trapezoids).

Such polyhedra are straightforward to translate into coherent boundary data: for top and bottom square, we choose anti-parallel normals. Without loss of generality we choose $\vec{e}_z$ for the top and $-\vec{e}_z$ for the bottom. We call the normals of the four ``side-faces'' $\vec{r}_l$. For this setup to be symmetric, we require that the dihedral angle between $\vec{e}_z$ and all $\vec{r}_l$ is equal and call it $\phi$. We derive the angle $\phi$ as a function of the spins $j_1$, $j_2$ and $k$ from the closure constraint:
\begin{equation}
	j_1 \vec{e}_z - j_2 \vec{e}_z + k \sum_{l=1}^4 \vec{r}_l = 0 \quad .
\end{equation}
We take the scalar product of this vector equation with $\vec{e}_z$ and demand $\vec{e}_z \cdot \vec{r}_l = \cos(\phi) \; \forall l$:
\begin{equation}
	j_1 - j_2 + 4 k \cos(\phi) = 0 \quad \iff \cos(\phi) = \frac{j_2 - j_1}{4 k} \quad .
\end{equation}
This equation readily implies restrictions on the spins for the angle $\phi$ to be well-defined, i.e. $-1 \leq \frac{j_2 - j_1}{4 k} \leq 1$.

Given the angle $\phi$, we fix the remaining normal vectors $\vec{r}_l$: we obtain them by a rotation of $\vec{e}_z$ around the $x$-axis by $\phi$ and a consecutive rotation around the $z$-axis by $(l-1) \frac{\pi}{2}$:
\begin{equation}
	|j, \vec{r}_l \rangle = e^{i (l-1) \frac{\pi}{2} J_z} e^{i \phi J_x} \triangleright |j,\vec{e}_z\rangle \quad ,
\end{equation}
where $l \in \{ 1,2,3,4 \}$. Eventually, the frustum intertwiner is defined as:
\begin{equation}
	\iota_{j_1,j_2,k} := \int_{\text{SU}(2)} dg \, g \triangleright |j,\vec{e}_z \rangle \, \otimes \, g \triangleright |j, -\vec{e}_z \rangle \bigotimes_{l=1}^4 g \triangleright |j, \vec{r}_l \rangle \quad .
\end{equation}

\begin{figure}
	\includegraphics[width = 0.5\textwidth]{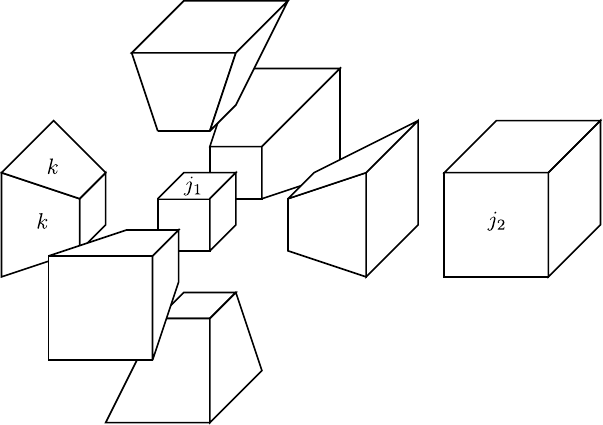}
	\caption{\label{fig:frustum_bdr}
		The boundary of a frustum, consisting of a small, a large cube and six identical frusta. The top and bottom faces of each frusta are glued to the small and large cube respectively, which are given by face areas $j_1$ and $j_2$ respectively. The only remaining degree of freedom is the size of the side faces of the frusta $k$. 
		}
\end{figure}

In the frustum case, the vertex amplitude is associated to a hyper-frustum, which is built from two cubes of different size\footnote{If the two cubes agree, the hyperfrustum simplifies to a (shape-matched) hypercuboid.} and six frusta that interpolate between them, see fig. \ref{fig:frustum_bdr}. The cubes have face areas $j_1$ and $j_2$ respectively, and the frusta are additionally described by the area of their side faces $k$. So, in total the hyper-frustum is described by three spins.

Compared to cuboids, frusta are more interesting from a physical point of view. On the one hand, they do allow for curved configurations, which leads to an amplitude with oscillatory behaviour. On the other hand, they describe a simple cosmological model: ``spatial'' slices are cubulated by identical cubes and can expand or contract under evolution via the frusta \cite{Bahr:2017eyi}.

\begin{figure}
	\includegraphics[width=0.5\textwidth]{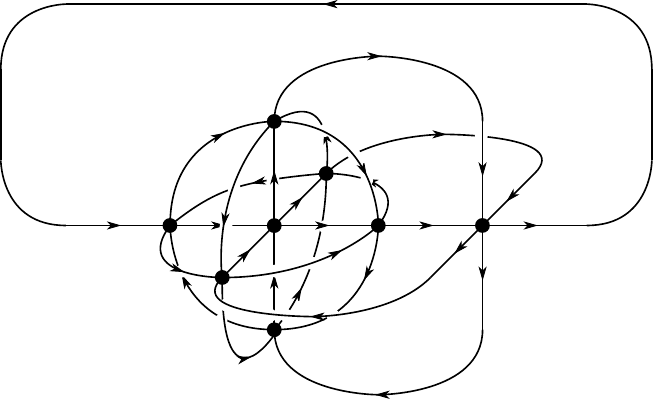}
	\caption{\label{fig:vertex_snw}
		Spin network graph corresponding to the vertex amplitude of a hypercubic 2-complex. Each node represents a six-valent intertwiner, which get contracted according to the combinatorics of the graph.
	}
\end{figure}

For both restricted models, the vertex amplitude is computed by contracting the intertwiners according to the vertex amplitude's spin network graph, see fig. \ref{fig:vertex_snw}. This graph is obtained by drawing a 2-sphere around the vertex in the 2-complex. Edges of the 2-complex pierce the sphere in a point, which become the nodes of the spin network, while the faces pierce the sphere along a line, which become the links of the spin network. These links come with a fiducial orientation: above we defined the intertwiners for all links outgoing. For opposite orientation, we choose the dual state peaked on $-\vec{n}$ for convenience. We discuss this and the semi-classical amplitude in the next section.

\subsection{Brief review of semi-classical amplitude}

Coherent intertwiners play a crucial role in deriving the asymptotic expansion of spin foam vertex amplitudes, which are also often referred to as their semi-classical limit. For coherent boundary data the vertex amplitude can be written as an integral over several copies of the symmetry group with an integrand given by inner products of coherent states:
\begin{equation}
	\mathcal{A}_v = \int_{\text{SU}(2)^N} \prod_{i=1}^N dg_i \prod_{a<b} \langle -\vec{n}_{ba} | g_b^{-1} g_a | \vec{n}_{ab} \rangle^{2 j_{ab}} \quad ,
\end{equation}
where $a$ and $b$ denote intertwiners, and pairs $ab$ the links connecting intertwiner $a$ to $b$. $|\vec{n}\rangle$ is the coherent state in the fundamental spin-$\frac{1}{2}$ representation and $\langle j, \vec{n}_1 | j, \vec{n}_2\rangle = \langle \vec{n}_1 | \vec{n}_2\rangle^{2 j}$ is a property of the maximal weight states.

To study the asymptotic expansion of the vertex amplitude in the Euclidean EPRL model we must consider $\text{SU}(2) \times \text{SU}(2)$ coherent states. As mentioned above, for $\gamma < 1$ the $\text{SU}(2)$ representation vector space $V_j$ is embedded into the unique Clebsch-Gordan decomposition of $V_{j^+} \otimes V_{j^-}$. For the maximum weight states and thus the coherent states this implies:
\begin{equation}
	|j,\vec{n}\rangle \mapsto |j^+,\vec{n}\rangle \otimes |j^-,\vec{n}\rangle \quad , \text{with } j^\pm = \frac{1}{2} (1 \pm \gamma) j \quad .
\end{equation}
The same applies for the intertwiners, which in the end can be written as the tensor product of two $\text{SU}(2)$ intertwiners with appropriate representations; the $\text{SU}(2)$ integration of the initial $\text{SU}(2)$ intertwiner can be absorbed due to the invariance of the Haar measure. Thus, we eventually find the expression for the coherent vertex amplitude, which factorizes into two expressions for ``$+$'' and ``$-$'' variables:
\begin{equation}
	\mathcal{A}_v = \mathcal{A}^+_v \mathcal{A}^-_v = \left ( \int_{\text{SU}(2)^N} \prod_{i=1}^N dg^+_i \prod_{a<b} \langle -\vec{n}_{ba} | (g^+_b)^{-1} g^+_a | \vec{n}_{ab} \rangle^{2 j^+_{ab}} \right ) \left ( \int_{\text{SU}(2)^N} \prod_{i=1}^N dg^-_i \prod_{a<b} \langle -\vec{n}_{ba} | (g^-_b)^{-1} g^-_a | \vec{n}_{ab} \rangle^{2 j^-_{ab}} \right ) \quad .
\end{equation}
Due to the factorisation of the amplitude, we can approximate each integral individually in a stationary phase approximation.

In a nutshell, the idea of a stationary phase approximation is that highly oscillatory integrals are dominated by stationary and critical points, for which the integral oscillates the least. For the amplitudes in question, we exponentiate the inner products of the coherent states and define the action:
\begin{equation}
	S^{\pm} := \sum_{a<b} 2 j_{ab} \ln \langle -\vec{n}_{ba} | (g^+_b)^{-1} g^+_a | \vec{n}_{ab} \rangle \quad .
\end{equation}
As we increase $j_{ab}$, e.g. by uniform scaling $\{j_{ab}\} \rightarrow \{\lambda j_{ab}\}$\footnote{The limit $\lambda \rightarrow \infty$ is often called semi-classical, because the areas in Planck units diverge. This corresponds to sending $l_p \rightarrow 0$.}, the integral oscillates faster and faster and is dominated by the contributions from critical points. These are determined by demanding the real part of the action to vanish as well as its variation with respect to the group elements $g_a$. The latter demands closure of the polyhedra, see eq. \eqref{eq:closure}, while the former determines how these polyhedra are glued together. If these conditions are not satisfied, the amplitude is exponentially suppressed. Configurations that are peaked on Regge geometries, e.g. corresponding to a Euclidean 4-simplex, possess two critical points, which differ by a phase given by the Regge action \cite{Regge:1961px}. Since these types of analyses were performed in great detail for various cases, we just briefly report the results relevant for this article and refer interested readers to the detailed literature \cite{Conrady:2008mk,Barrett2009a} (here for the Euclidean 4-simplex amplitude of the EPRL model).

\subsubsection{Asymptotic hypercuboid vertex amplitude}

For cuboid intertwiners, the following semi-classical formula was derived in \cite{Bahr:2015gxa}:
\begin{equation}
	\mathcal{A}^{\pm}_v = \left ( \frac{1 \pm \gamma}{2} \right )^{\frac{21}{2}} \mathcal{B}_v \quad ,
\end{equation}
where
\begin{equation}
	\mathcal{B}_v = \left (\frac{2}{16 \pi^2} \right )^7  (2 \pi)^{\frac{21}{2}} \left ( \frac{1}{\sqrt{\det -H}} + \text{c.c.} \right ) \quad .
\end{equation}
$\det -H$ denotes the determinant of the Hessian matrix, which is a rational function of the six spins characterizing the vertex amplitude. The pre-factors denote the integration volume of $\text{SU}(2)$ times the multiplicity of critical points, here for seven integrations, and the usual stationary phase pre-factor for integrating over 21 variables. Under uniform scaling of all spins $j \rightarrow \lambda j$, $\det H \rightarrow \lambda^{21} \det H$, such that the amplitude scales as $\lambda^{-\frac{21}{2}}$. Note that this amplitude is non-oscillatory and positive, i.e. the Regge action associated to hypercuboids vanishes identically independent of the spins. 

\subsubsection{Asymptotic frustum vertex amplitude}

For frustum intertwiners the formula is of a similar form, but more intricate \cite{Bahr:2017eyi}. Again the vertex amplitude factorises into two amplitudes for ``$+$'' and ``$-$'' variables:
\begin{equation}
	\mathcal{A}^{\pm}_v = \left ( \frac{1 \pm \gamma}{2} \right )^{\frac{21}{2}} \mathcal{B}^{\pm}_v \quad ,
\end{equation}
where the overall scaling comes again from the determinant of the Hessian matrix. In contrast to the cuboid case, the Regge action does not vanish, thus the amplitude $\mathcal{B}_v$ retains a dependence on the Immirzi parameter $\gamma$.
\begin{equation}
	\mathcal{B}^\pm_v = \left (\frac{1}{8 \pi^2} \right )^7 (2 \pi)^{\frac{21}{2}} \left ( \frac{e^{i (1 \pm \gamma) S_R}}{\sqrt{\det -H}} + \text{c.c.} \right ) \quad ,
\end{equation}
where $S_R$ denotes the Regge action of a hyper-frustum:
\begin{equation}
	S_R := 6 j_1 \left ( \frac{\pi}{2} - \Theta \right ) + 6 j_2 \left ( \frac{\pi}{2} - \Theta' \right ) + 12 k \left ( \frac{\pi}{2} - \Theta'' \right ) \quad ,
\end{equation}
where $j_1$, $j_2$ and $k$ denote the $\text{SU}(2)$ spins. The exterior dihedral angles are functions of the angle $\phi$ and thus of the spins.
\begin{align}
	\Theta & = \theta \\
	\Theta' & = \pi - \theta \\
	\Theta'' & = \arccos( \cos^2 \theta) \quad ,
\end{align}
where $ \theta = \arccos \frac{1}{\tan \phi}$. These exterior dihedral angles are located in the hyper-frustum as follows: $\Theta$ and $\Theta'$ are associated to the faces that are shared by the initial final cube and the frusta. $\Theta''$ is belongs to faces shared between frusta. When inserting the formula for $\mathcal{B}^\pm$ into $\mathcal{A}_v$ one obtains an analogous result to \cite{Barrett2009a}: one term corresponds to the cosine of $\gamma S_R$, whereas the remaining two terms also oscillate with the Regge action, yet without $\gamma$ dependence.

Note also that the 4D dihedral angles enforce more strict constraints on the labels $j_1$, $j_2$ and $k$ than the angle $\phi$ already does. For the angle $\theta$ to be well-defined, we must demand \cite{Bahr:2017eyi}:
\begin{equation}
	-\frac{1}{\sqrt{2}} \leq \frac{j_2 - j_1}{4 k}  \leq \frac{1}{\sqrt{2}} \quad .
\end{equation}
Thus, there exist spin configurations $\{j_1,j_2,k\}$ for which the intertwiners are well-defined, but the semi-classical amplitude is not, i.e. no critical points exist and the amplitude is exponentially suppressed for large spins. This can be checked numerically in principle.

Since the vertex amplitude (for $\gamma < 1$) factorises essentially into the product of the amplitudes $\mathcal{B}^\pm_v$, we will focus in this article on the numerical evaluation of $\mathcal{B}_v$ in both cuboid in frustum cases. If we find a good agreement (or convergence) of the full expression to its semi-classical approximation, these findings automatically generalize to the full amplitude $\mathcal{A}_v$, and we can save vital computational time.

In the following section, we discuss the numerical setup to calculate the amplitude $\mathcal{B}_v$ and the numerical costs.

\section{Numerical evaluation of the vertex amplitude} \label{sec:numerical}

The numerical implementation of the calculation of the vertex amplitude is split into two parts: firstly, we compute the intertwiners relevant for computing a particular vertex amplitude. This is done by defining the tensor product of coherent $\text{SU}(2)$ states and explicit group averaging. Secondly, these intertwiners are contracted with respect to boundary graph of the vertex, i.e. the respective magnetic indices of the intertwiners are identified and summed over.

\subsection{Definition of coherent intertwiners}

An essential ingredient for the definition of the coherent intertwiners are the Wigner $D$-matrices, which encode the action of the group on the states. In this article, we define them in the ``$z-x-z$'' convention of Euler angles $\alpha$, $\beta$ and $\gamma$ (not to be confused with the Immirzi parameter):
\begin{equation}
	D^j_{m' m}(\alpha,\beta,\gamma) := \langle j,m' | e^{-i \alpha J_z} e^{-i \beta J_x} e^{-i \gamma J_z} | j, m  \rangle] = e^{-i m' \alpha} \, d^j_{m' m}(\beta) \, e^{-i m \gamma} \quad ,
\end{equation}
where $j$ labels the irreducible representation, and $m',m$ are magnetic indices labeling the basis elements of the representation vector space $V_j$. $d^j(\beta)$ denotes the small Wigner $d$-matrix, given by the matrix elements of the exponential of the $J_x$ generator, with $\alpha \in [0,2\pi]$, $\beta \in [0,\pi]$ and $\gamma \in [0,4\pi]$.

The coherent intertwiners are defined as group averaged tensor products of $\text{SU}(2)$ coherent states, which are peaked on a direction $\vec{n} \in S^2$, i.e. they diagonalize the generator of rotations $J_{\vec{n}}$ associated with this directions. We define these states by starting from the same states ($|j,-j\rangle$ for cuboids and $|j,j\rangle$ for frusta) and act on each with a specific Wigner $D$-matrix rotating the state into the desired direction. In the following we specify the angles for each 3D normal vector in both cuboid and frustum cases\footnote{$\text{SU}(2)$ coherent states are defined up to a phase. This also implies that the vertex amplitude is defined up to a global phase.}.

Each intertwiner has three ingoing and three outgoing links. In anticipation of the contraction, we make the following assignment of states:
\begin{align}
	|j_{ab},\vec{n}_{ab}\rangle & \quad \text{outgoing link } a \rightarrow b \\
	|j_{ab},-\vec{n}_{ab}\rangle^\dagger \equiv \langle j_{ab}, -\vec{n}_{ab} | & \quad \text{ingoing link } a \leftarrow b
\end{align}
With this convention, the vertex amplitude is straightforwardly given by identifying the indices of the intertwiners and summing over them.

Eventually we define the intertwiner by group averaging, i.e. we act with the same $g$ on all states and integrate $g$ over $\text{SU}(2)$. In practice, we parametrize $g$ by the three Euler angles $\alpha$, $\beta$ and $\gamma$ introduced above and act with $D(g(\alpha,\beta,\gamma))$ on ``outgoing'' states and $D^\dagger(g(\alpha,\beta,\gamma))$ on the dual ``ingoing'' states. In the parametrization, the $\text{SU}(2)$ Haar measure reads:
\begin{equation}
	dg = \frac{1}{16 \pi^2} \sin(\beta) d\alpha \, d\beta \, d\gamma \quad .
\end{equation}

Combining all these ingredients, the intertwiners are defined as follows:

\begin{align}
	\iota_{\{j_i\}} & = \int_{\text{SU(2)}} dg \; \bigotimes_{i \text{ outgoing}} D^{j_i}(g) |j_i,\vec{n}_i\rangle \; \bigotimes_{i \text{ ingoing}} \left( D^{j_i}(g) |j_i,-\vec{n}_i\rangle \right)^\dagger \nonumber \\
	& = \int_{\text{SU(2)}} dg \; \bigotimes_{i \text{ outgoing}} D^{j_i}(g) D^{j_i}(h) |j_i,\vec{e}_z\rangle \; \bigotimes_{i \text{ ingoing}} \left( D^{j_i}(g)  D^{j_i}(h_i)|j_i,\vec{e}_z\rangle \right)^\dagger \label{eq:intertwiner}
\end{align}
Here $h_i \in \text{SU}(2)$ denotes the group element rotating $\vec{e}_z \rightarrow \vec{n}_i$, which in turn are again parametrised by angles $\alpha_i$, $\beta_i$ and $\gamma_i$. We will specify these angles below in both the cuboid as well as the frustum case.

Numerically we define the intertwiners as an array with six indices, as many as the node has links. In appendix \ref{app:notation} we define the convention how to enumerate the intertwiners and their indices. Each component is explicitly defined following eq. \eqref{eq:intertwiner} as a three-dimensional integral in $\alpha$, $\beta$ and $\gamma$. We compute each component numerically using the \texttt{Cuba}\footnote{\url{https://github.com/giordano/Cuba.jl}} package \cite{Hahn:2004fe,Hahn:2014fua} in the programming language \texttt{Julia}, more precisely the algorithm \texttt{cuhre}.
Since each component of the intertwiner is independent, its computation can be straightforwardly parallelized across multiple cores and nodes.

In the next two subsections we briefly state the definitions of cuboid and frustum intertwiner explicitly via the angles $\alpha_i$, $\beta_i$ and $\gamma_i$ chosen to rotate $\vec{e}_z \rightarrow \vec{n}_i$.

\subsubsection{Boundary data of cuboid intertwiners}

The outward pointing normals to a cuboid are easy to parametrise: the normal to one face is orthogonal to all the normal vectors of its four adjacent faces and it is antiparallel to the normal assigned to the opposite face. Hence, we choose the Cartesian basis vectors as the normal vectors to the faces of the cuboids. The respective Euler angles are given in table \ref{tab:cuboid_angles}\footnote{We use the so-called $z-x-z$ notation for the Wigner matrices.}.
\begin{table}[h!]
	\centering
	\begin{tabular}{|c | c | c | c |}
		\hline
		$\; \vec{n} \;$ & $\alpha$ & $\beta$ & $\gamma$ \\
		\hline
		$\; \vec{e}_{x} \;$ & $\; - \frac{\pi}{2} \;$ & $\frac{\pi}{2}$ & $ \frac{\pi}{2}$ \\
		\hline
		$\; -\vec{e}_{x} \;$  & $\; - \frac{\pi}{2} \;$ & $ \; - \frac{\pi}{2} \;$ & $\; \frac{\pi}{2} \;$ \\
		\hline
		$\vec{e}_{y}$ & $0$ & $ - \frac{\pi}{2}$ & $ 0$ \\
		\hline
		$ \; -\vec{e}_{y} \;$ & $ 0$ & $\frac{\pi}{2}$ & $0$ \\
		\hline
		$\vec{e}_{z}$ &  $0$ & $ 0$ & $ 0$ \\
		\hline
		$\; -\vec{e}_{z} \;$ & $0$ & $\pi$ & $\pi$ \\
		\hline
	\end{tabular}
\caption{\label{tab:cuboid_angles}
	Table of Euler angles parametrizing the boundary states for cuboid intertwiners.
}
\end{table}

\subsubsection{Boundary data of frustum intertwiners}

Two types of intertwiners are required for the frustum case: cube intertwiners and frustum intertwiners. The former are explained in the previous subsection; here we focus on the frustum case. In particular, we must define the Euler angles for the normal $\vec{r}_l$. In addition to these states, we must also define the angels for the dual states labeled by $-\vec{r}_l$. The angles\footnote{These choices for the angles give an amplitude with a global phase $\exp{-i (j_2-j_1) \frac{\pi}{2}}$. We correct for this such that the final amplitude is real.} are summarized in tab. \ref{tab:angles_frustum}\footnote{Attentive readers will notice the slight difference in definition of rotations. This is due to a slight difference in the implementation of the codes for cuboids and frusta. For cuboids we act on maximal states $|j,-j\rangle$, whereas we act on $|j,j\rangle$ for frusta. While this does not affect the results, we mention it here for transparency.}.

\begin{table}[h!]
	\centering
	\begin{tabular}{|c | c | c | c |}
		\hline
		$\; \vec{n} \;$ & $\alpha$ & $\beta$ & $\gamma$ \\
		\hline
		$\; \vec{e}_{r_1} \;$ & $\;- \frac{\pi}{2} \;$ & $\;-\phi \;$ & $\; \frac{\pi}{2} \;$ \\
		\hline
		$\; -\vec{e}_{r_1} \;$ & $\; - \frac{\pi}{2} \;$ & $\; \pi - \phi \;$ & $\; \frac{\pi}{2} \;$ \\
		\hline
		$\; \vec{e}_{r_2} \;$ & $\;  0 \;$ & $\;  - \phi \;$ & $\;  0 \; $ \\
		\hline
		$\; -\vec{e}_{r_2} \;$ & $\; 0 \;$ & $\; \pi - \phi \;$ & $\;  0 $ \\
		\hline
		$\; \vec{e}_{r_3} \;$ & $\; - \frac{\pi}{2} \;$ & $\; \phi \;$ & $\; \frac{\pi}{2} \;$ \\
		\hline
		$\; -\vec{e}_{r_3} \;$ & $\; - \frac{\pi}{2} \;$ & $\; -\pi + \phi \;$ & $\; \frac{\pi}{2} \;$ \\
		\hline
		$ \; \vec{e}_{r_4} \;$ & $\;  0 \;$ & $\;  \phi \;$ & $\; 0 \;$ \\
		\hline
		$\; -\vec{e}_{r_4} \;$ & $\;  0 \;$ & $\;  - \pi + \phi \;$ & $\;  0 \;$ \\
		\hline
		$\; \vec{e}_{z} \;$ & $\; 0 \;$ & $\; 0 \;$ & $\; 0 \;$ \\
		\hline 
		$\; -\vec{e}_{z} \;$ & $\; 0 \;$ & $\; \pi \;$ & $\; - \pi \;$ \\
		\hline
	\end{tabular}
	\caption{ \label{tab:angles_frustum}
		Table of Euler angles parametrizing boundary states of frustum intertiners.
	}
\end{table}

\subsection{Definition of vertex amplitude}

After defining the boundary data for each intertwiner and performing the $\text{SU}(2)$ integration for each, we compute the vertex amplitude by contracting the indices of the intertwiners according to the vertex graph in fig. \ref{fig:vertex_snw}. In the code, we concretely define this contraction by picking a notation for each intertwiner, identifying the shared magnetic indices and summing over them. This is straightforwardly possible since we have defined the intertwiners with the subsequent contraction in mind. In appendix \ref{app:notation} we explain the notation and give the explicit expression of the vertex amplitude in eq. \eqref{eq:vertex_amplitude_code}.

\subsection{Optimization and (scaling of) numerical costs}

Previous attempts to numerically calculate the full spin foam vertex amplitude focused on the vertex amplitudes of 4-simplices. 4-simplices are the simplest discrete structures to span a 4D space (or space-time in the Lorentzian setting), and thus require less data to be fully specified compared to higher-valent building blocks. By considering cuboids and frusta we inadvertently have to face larger numerical costs. Take the computation of a single 4-valent versus a 6-valent intertwiner, where all faces carry spin $j$. Naively, a 4-valent intertwiner has $(2j+1)^4$ components while the 6-valent intertwiner has $(2j+1)^6$ components. Similarly, contracting the intertwiners requires us to sum over ten magnetic indices for a 4-simplex, while we sum over 24 indices for hypercuboids / -frusta. Due to this scaling behavior, optimizations are vital. In the following we explain the optimizations we use in this work. Optimizations beyond these are possible, yet require additional work to be implemented for higher-valent spin foam vertex amplitudes. We discuss those in more detail in the discussion in section \ref{sec:discussion}.

Using the symmetries of intertwiners, we can significantly improve the scaling behavior sketched above. As $\text{SU}(2)$-invariant tensors, intertwiners only span a subspace of the full tensor product space of representation vector spaces. Hence, there are components of intertwiners that always vanish due to symmetry, which need not be explicitly computed nor summed over. Clearly, implementing these symmetries leads to lower computational costs. In case of six-valent intertwiners, where we assume (without loss of generality) its first three indices as outgoing and its latter three as ingoing, this condition is given by:
\begin{equation}
m_1 + m_2 + m_3 - m_4 - m_5 - m_6 = 0 \quad ,	
\end{equation}
where $m_i$ denote the magnetic indices for each link of the intertwiner. With this condition we determine which components of the intertwiners always vanish and thus can be safely ignored. In practice, we implement this condition in two different ways. When computing the intertwiners, we explicitly check whether $\{m_i\}$ satisfy this condition before and store the allowed configurations. Then, we compute only the allowed components explicitly and set all other ones to zero. For the contraction, we take seven intertwiners and fix one of their indices as a function of the remaining ones. Additionally, we check whether this solution is permitted by the representation, i.e. $-j_i \leq m_i \leq j_i$. After this condition is implemented for seven intertwiners, it is automatically satisfied for the final remaining one. This optimization greatly reduces the numerical costs, which allows us to explore larger representations. However, increasing the representation labels still leads to a rapid growth of costs, which can be partially compensated by parallelizing the code.

\subsubsection*{Parallelization}

Our code is split into two parts: firstly, we compute the intertwiners component by component performing a three-dimensional integration over $\text{SU}(2)$ for each component. The final tensor is then stored in a text file; that way we compute each intertwiner only once and reuse it if it is part of a different vertex amplitude. Secondly, we compute the contraction of the intertwiners: we read in the intertwiners from the text files and contract them in a \verb|for|-loop over the non-fixed magnetic indices.

Both parts of the code are parallelizable: when calculating an intertwiner, each of its components is entirely independent and thus can be distributed across multiple cores (and nodes). We used \verb|Julia|'s included \verb|Distributed| package to do so. For the contraction of the intertwiners to obtain the vertex amplitude, we split the set of seventeen \verb|for| loops into two. The first one runs over the five non-fixed magnetic indices of the first intertwiner, which is parallelized using \verb|Julia|'s built-in \verb|multi-threading| functionality. The task to compute the summands of the vertex amplitude is thus split across multiple threads and cores. To synchronize the results we use the \verb|atomic| sum routine in \verb|Julia|.

\begin{figure}
	\includegraphics[width=0.5\textwidth]{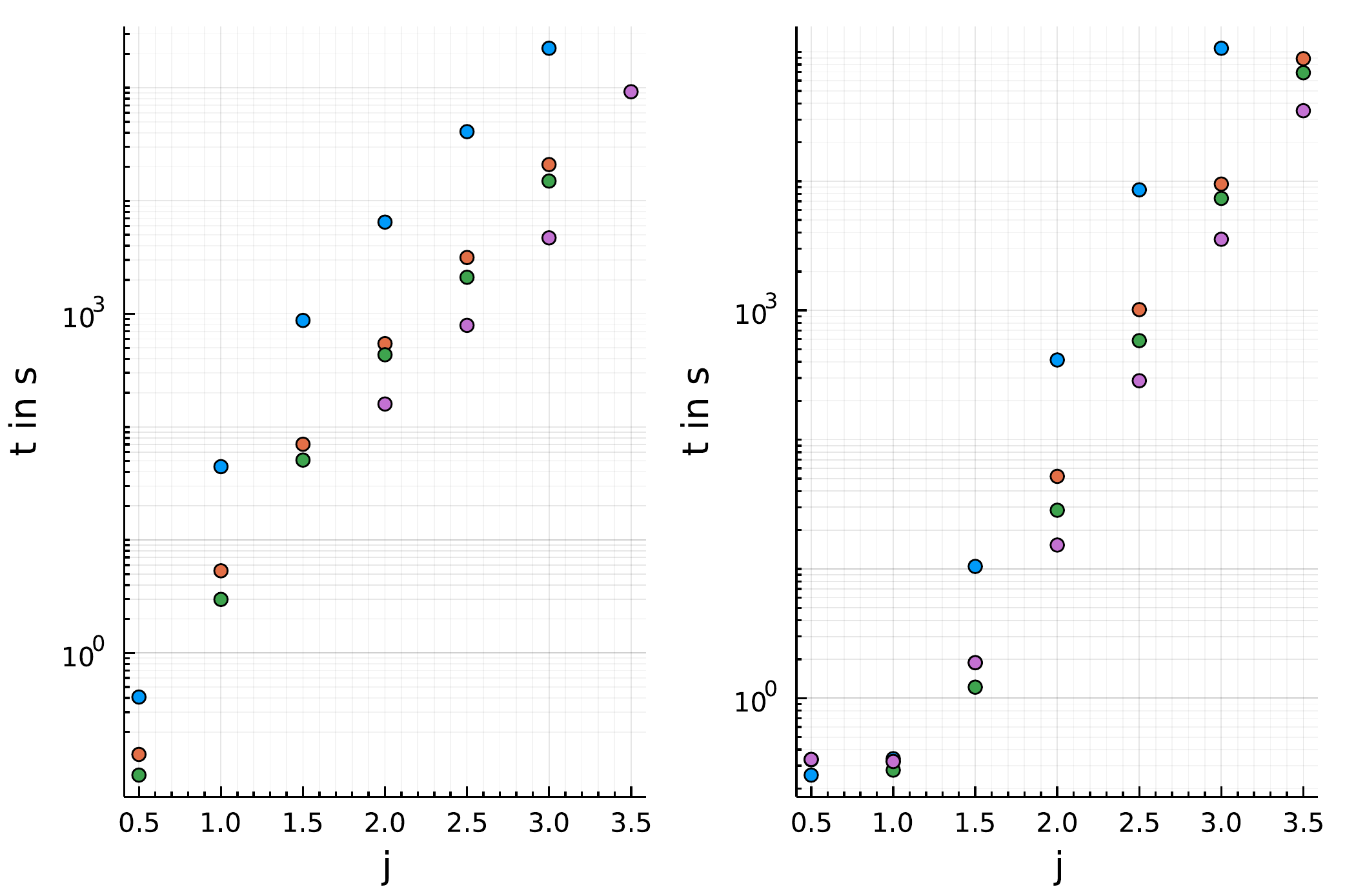}
	\caption{\label{fig:compute_time}
	\textit{Left:} Time to compute one cube intertwiner with all spins equal to $j$. Colors mark different number of cores: one core (blue), 16 cores (orange), 32 cores (green) and 64 cores on different nodes (purple). Points omitted for 64 cores and small spins.
	\textit{Right:} Time to compute the hypercube vertex amplitude for all spins $j$. Colors show different number of cores used: one core (blue), 16 cores (orange), 32 cores (green) and 72 cores (purple).
}
\end{figure}

Despite these efforts, both parts of the algorithm quickly require substantial computational resources and time. In fig.~\ref{fig:compute_time}, we show the time it takes to compute one cube intertwiner with all spins $j$ and the time it takes to compute the contraction of eight cube intertwiners with all spins $j$, while using different number of cores on the Ara cluster in Jena\footnote{Each node is equipped with two Intel Xeon Gold 6140 processors (18 Core 2,3 Ghz, Skylake architecture).}. In both cases, the numerical costs scale exponentially as we increase the spin $j$ despite the optimizations mentioned before. Parallelization substantially reduces the computational time, but cannot overcome the exponential growth.
Thus, the rapidly growing numerical costs imply that we cannot increase the representations labels indefinitely. Instead, we will focus on the cases that are feasibly accessible and compare the exact vertex amplitudes to the semi-classical approximations for the cuboid \cite{Bahr:2015gxa} and frustum models \cite{Bahr:2017eyi}. Our code is publicly available\footnote{\url{https://github.com/CourtA96/SpinfoamAmplitudesOfCuboidsAndFrustra}} and we also provide the data from our simulations\footnote{\url{https://doi.org/10.5281/zenodo.6006163}}.

\section{Results} \label{sec:results}

In this section we present our results for the cuboid and frusta vertex amplitudes and compare them to their semi-classical counterparts, the first 4D result for higher valent spin foam vertex amplitudes.
The semi-classical amplitude for coherent boundary data is expected to be valid in a limit where the representation labels are uniformly scaled up to be large, i.e. $\mathcal{A}_v(\{\lambda j_i\})$ for $\lambda \gg 1$. 
In our results, we explore this limit as best as possible, but are limited by the exponentially growing computational costs. While we see evidence that the full and semi-classical amplitude converge as we increase the spins, we cannot reach sufficiently high spins to show it explicitly, in particular in the frustum case.

\subsubsection{Beyond uniform scaling}

A flat Euclidean 4-simplex is uniquely determined (up to rotations, translations...) by its ten edge lengths, which determine the 4D dihedral angles and areas of triangles and thus its Regge action. Furthermore, we can translate this 4-simplex into coherent spin foam boundary data, i.e. areas and 3D normals. Conversely, if we only know the 4D dihedral angles of the 4-simplex, this fixes the 4-simplex up to scale, i.e. we exactly know its shape but not its size. In terms of spin foam boundary data, this fixes the 3D normals and the relative areas of the triangles; by universally scaling all spins (areas), we obtain a scaled 4-simplex of the same shape. Thus, universal scaling is straightforward to study, since we do not need to change the normal vectors encoded in the boundary states.

However, if we want to change individual spins / areas in a 4-simplex, we inevitably change the shape of the corresponding 4-simplex and must compute the associated 3D normals to find a non-suppressed vertex amplitude. This reveals an unexpected advantage of cuboid and frustum spin foams; due to the high degree of symmetry we can quite freely change single representations. In the cuboid case, the 3D normals are fixed by definition and all assignments of spins satisfy both $\text{SU}(2)$ coupling rules and lead to a non-suppressed vertex amplitude in the semi-classical limit. For frusta, the case is a bit more intricate: $\text{SU}(2)$ coupling rules may be violated and depending on the spin assignments, the 3D normals must be adapted. Fortunately, the latter are given as simple functions of the spins and thus straightforwardly realized.

This flexibility gives us the opportunity to kill two birds with one stone: one the one hand, it gives us the unique opportunity to study spin foam vertex amplitudes with highly different spins, e.g. some small and some large. In these cases, it is not known if and when the semi-classical approximation becomes accurate. One the other hand, we can partially tame the numerical costs of computing intertwiners and the vertex amplitude and still explore interesting and unknown regions of the theory.

\subsection{Cuboid results}

Let us begin by presenting the results for the vertex amplitude for cuboid intertwiners. Recall that the semi-classical approximation shows no oscillatory behavior, is positive and determined by the scaling behavior from a stationary phase approximation. In general, we expect that the semi-classical amplitude cannot be trusted for small spins, e.g. it generically diverges since it is written as an expansion in $j^{-1}$; hence we expect the full amplitude to have a different scaling behavior in the quantum regime. Whether the full amplitude also shows no oscillatory behavior is a priori not clear.

\begin{figure}
	\includegraphics[width=0.6\textwidth]{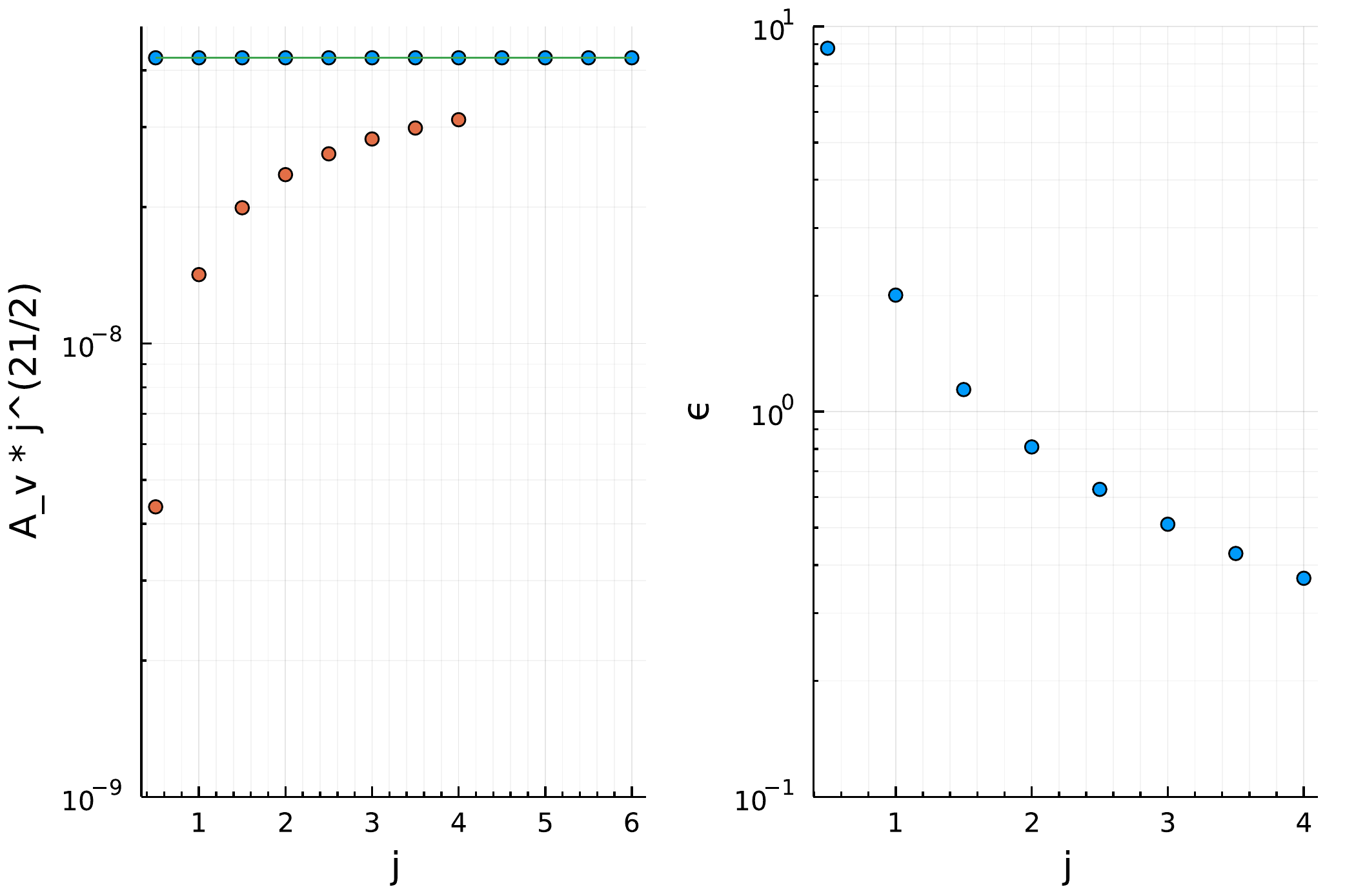}
	\caption{ \label{fig:cube_plots}
		\textit{Left side:} Semi-classical (blue) and full (orange) vertex amplitude $\mathcal{A}_v$ for all $j_i = j$ multiplied with $j^{\frac{21}{2}}$ with logarithmic scale. Green line is the semi-classical amplitude for continuous values of $j$. \textit{Right side:} Relative error $\epsilon$ of semi-classical ampltitude to the full one in logarithmic scale.
	}
\end{figure}

\begin{figure}
	\includegraphics[width=0.6\textwidth]{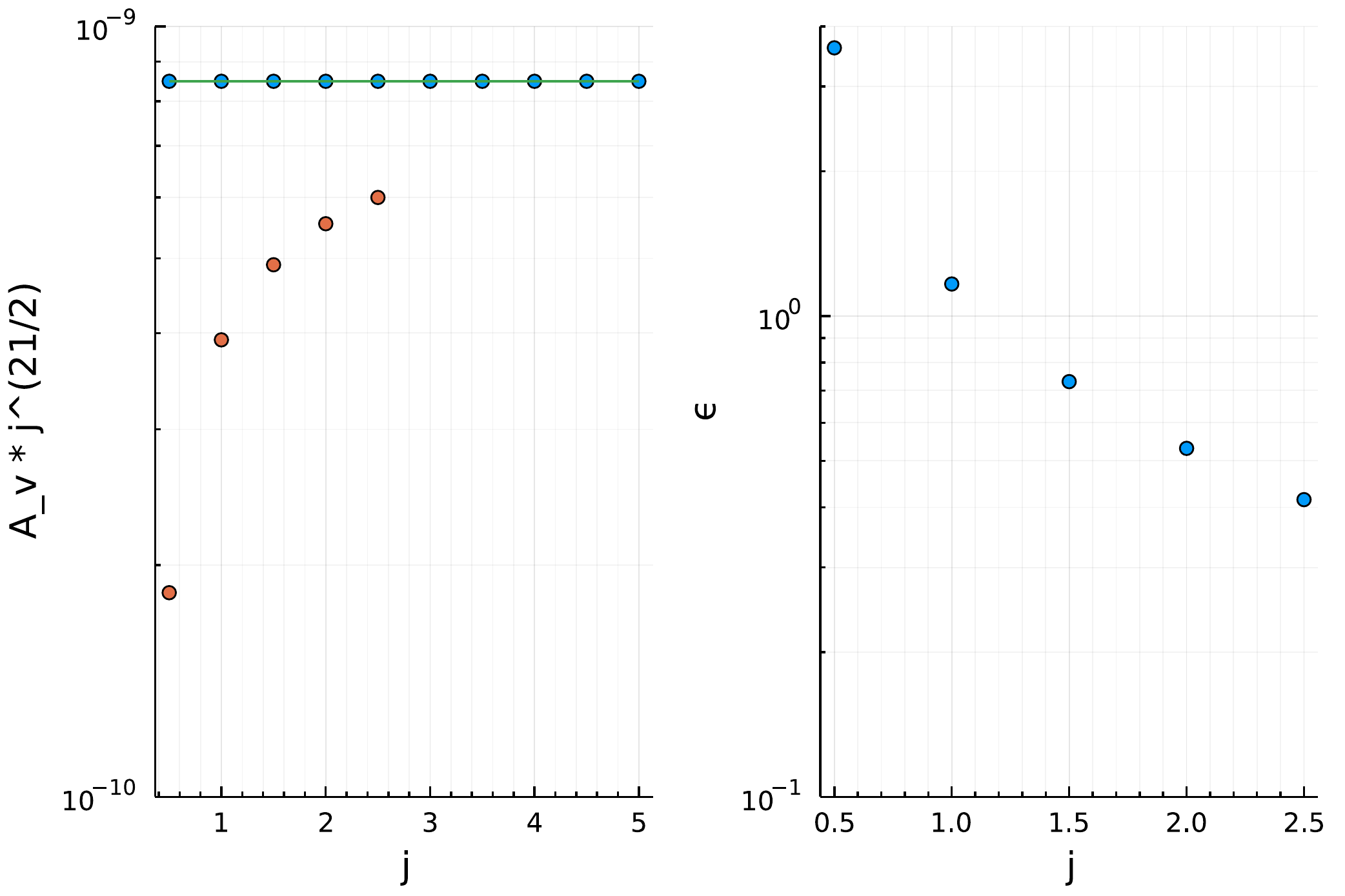}
	\caption{ \label{fig:cube_scale_plots}
		\textit{Left side:} Semi-classical (blue) and full (orange) vertex amplitude $\mathcal{A}_v$ for $j_1 = j_2 = j_3 = j$ and $j_4 = j_5 = j_6 = 2j$ multiplied with $j^{\frac{21}{2}}$ with logarithmic scale. Green line is the semi-classical amplitude for continuous values of $j$. \textit{Right side:} Relative error $\epsilon$ of semi-classical ampltitude to the full one in logarithmic scale.
	}
\end{figure}

In general, our numerical results confirm our expectations. The semi-classical amplitude overestimates the amplitude at small spins, while they quickly approach each other if all spins become large, i.e. if one is approaching the regime of validity of the stationary phase approximation. This can be seen for the vertex amplitude where all spins are chosen equal, see fig. \ref{fig:cube_plots}\footnote{The amplitude is multiplied with the semi-classical scaling behaviour.}. While the difference is large for all spins $j = \frac{1}{2}$, both amplitudes are fairly close already at spins $j = 4$. Its relative error $\epsilon = \frac{| \mathcal{A}^\text{full}_v - \mathcal{A}^\text{s.c.}_v |}{\mathcal{A}^\text{full}_v}$, it drops from $877 \%$ to $37\%$, see also fig. \ref{fig:cube_plots}. Additionally, the full amplitude shows no oscillatory behavior and is strictly positive. 
Results for spins $j_1, j_2, j_3 = j$ and $j_4,j_5,j_6 = 2j$ are similar, see fig. \ref{fig:cube_scale_plots}. Here the relative error drops to $42 \%$ for $j = 2.5$.

\begin{figure}
	\includegraphics[width=0.6\textwidth]{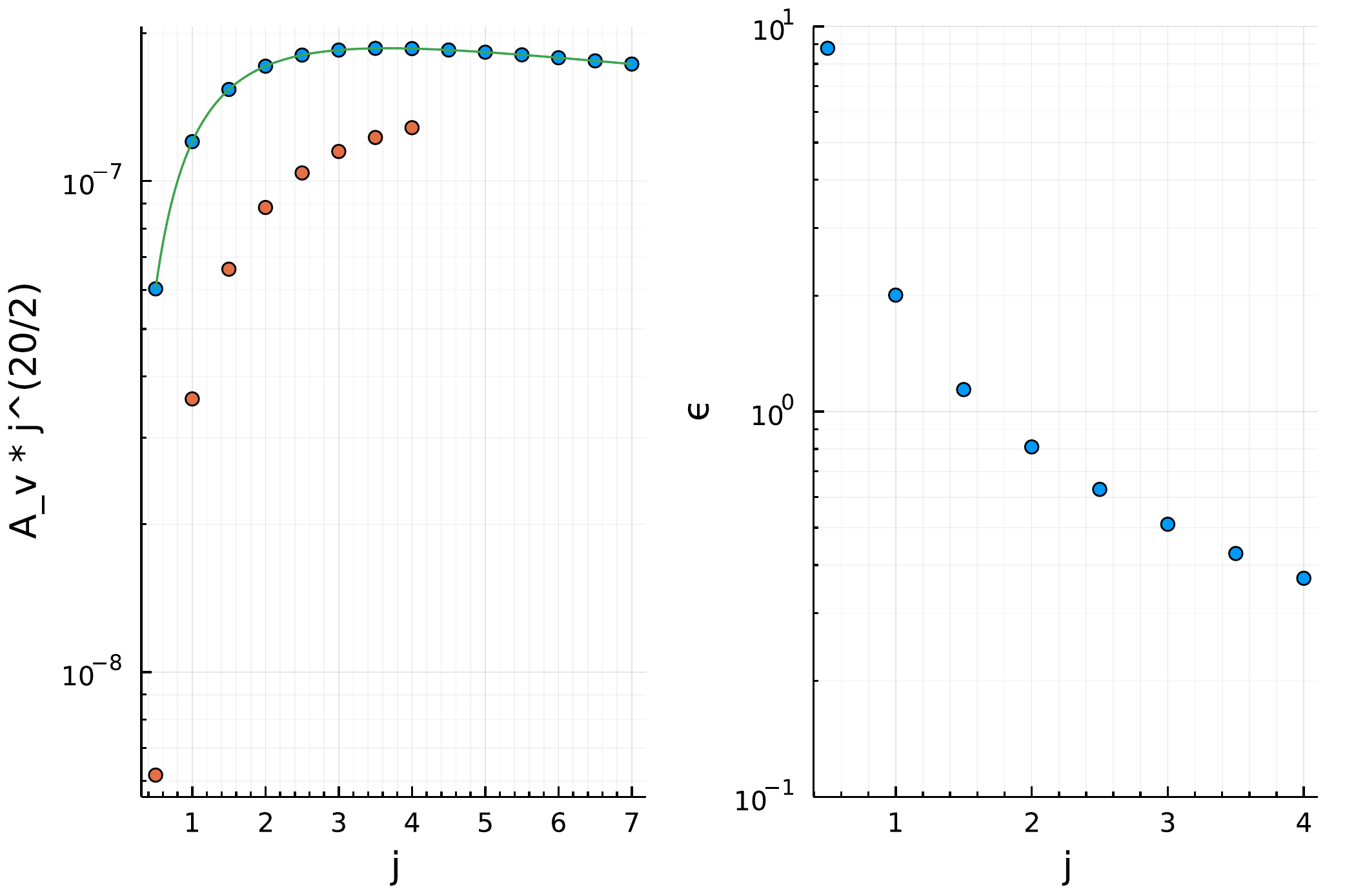}
	\caption{ \label{fig:plot_one_small}
		\textit{Left side:} Semi-classical (blue) and full vertex amplitude (orange) $\mathcal{A}_v$ for $j_1 =0.5$ and all remaining $j_i = j$ multiplied with $j^{\frac{20}{2}}$ with logarithmic scale. Green line show the semi-classical amplitude for continuous values of $j$. \textit{Right side:} Relative error $\epsilon$ of the semi-classical ampltitude to the full one in logarithmic scale.
	}
\end{figure}

Beyond the uniform scaling of all spins, we explore cases where we keep a few spins fixed and small, while gradually increasing the remaining ones. 
The first example is to keep one spin, e.g. $j_1$ small, while increasing all remaining ones. This case corresponds to a hypercuboid with torsion \cite{Bahr:2015gxa}; it is built from cuboids and cubes whose faces' shapes do not match. Despite this fact, the results are comparable to the uniform scaling of a hypercube, see fig. \ref{fig:plot_one_small}: the full and semi-classical amplitude quickly approach each other, the relative error drops to $45 \%$ at $j=4$ compared to $36 \%$ in the hypercube case. Thus, we think it is plausible that in this case the semi-classical amplitude becomes valid at sufficiently large spins despite the fact that one spin remains small.

\begin{figure}
	\includegraphics[width=0.6\textwidth]{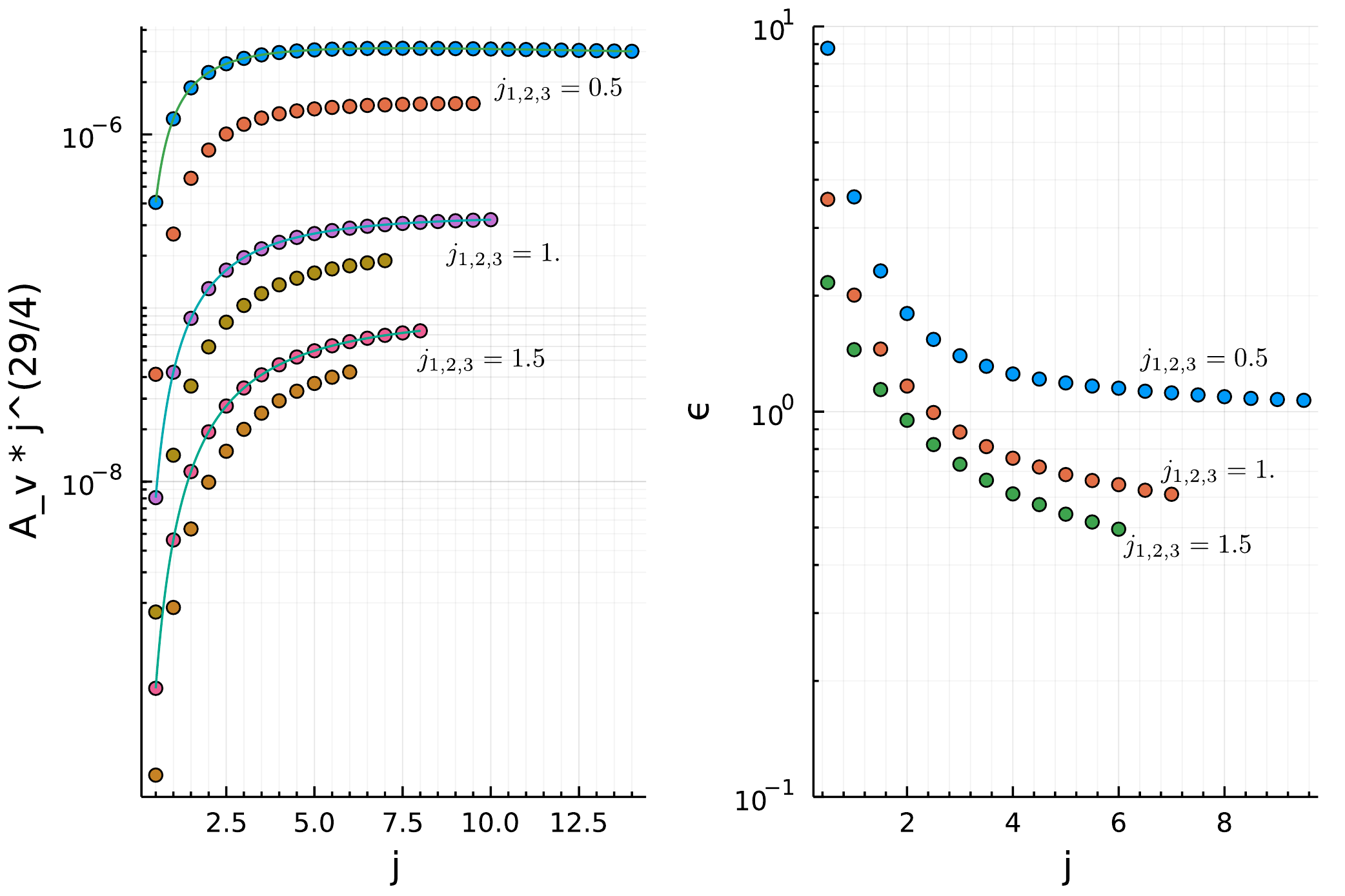}
	\caption{ \label{fig:plot_small_cubes}
		\textit{Left side:} Semi-classical and full vertex amplitude $\mathcal{A}_v$ for $j_1 = j_2 = j_3 = 0.5, \, 1, \, 1.5$ and $j_4 = j_5 = j_6 = j$ multiplied with $j^{\frac{29}{4}}$ with logarithmic scale. The plots at the top are for $j_1 = j_2 = j_3 = 0.5$ followed by $j_1 = j_2 = j_3 = 1$ and $j_1 = j_2 = j_3 = 1.5$. The semi-classical amplitude is always larger than the full one. Solid lines show the semi-classical amplitudes for continuous values of $j$. \textit{Right side:} Relative error $\epsilon$ of semi-classical ampltitudes to the full one in logarithmic scale, again $j_1 = j_2 = j_3 = 0.5$ followed by $j_1 = j_2 = j_3 = 1$ and $j_1 = j_2 = j_3 = 1.5$.
	}
\end{figure}

The next example is to fix three spins $j_1,j_2,j_3 = j$ at some small value $j$ and gradually increase the remaining ones. These are shape-matching configurations: we have two small cubes which are connected by cuboids, whose side rectangle become larger and larger. In fig. \ref{fig:plot_small_cubes} we summarize three plots for $j= \frac{1}{2}, 1, \frac{3}{2}$: As in the previous examples, we do observe a convergence between the full and semi-classical amplitude, but significantly slower despite the fact that we can explore larger spins. For $j=\frac{1}{2}$ the relative error remains around $100\%$ and only slowly decreases, while it rapidly improves to above $60\%$ for $j=1$ and $50\%$ for $j=\frac{3}{2}$. The two latter cases suggest that a semi-classical regime could be reached even if some spins remain small as long as other spins become large.

\begin{figure}
	\includegraphics[width=0.6\textwidth]{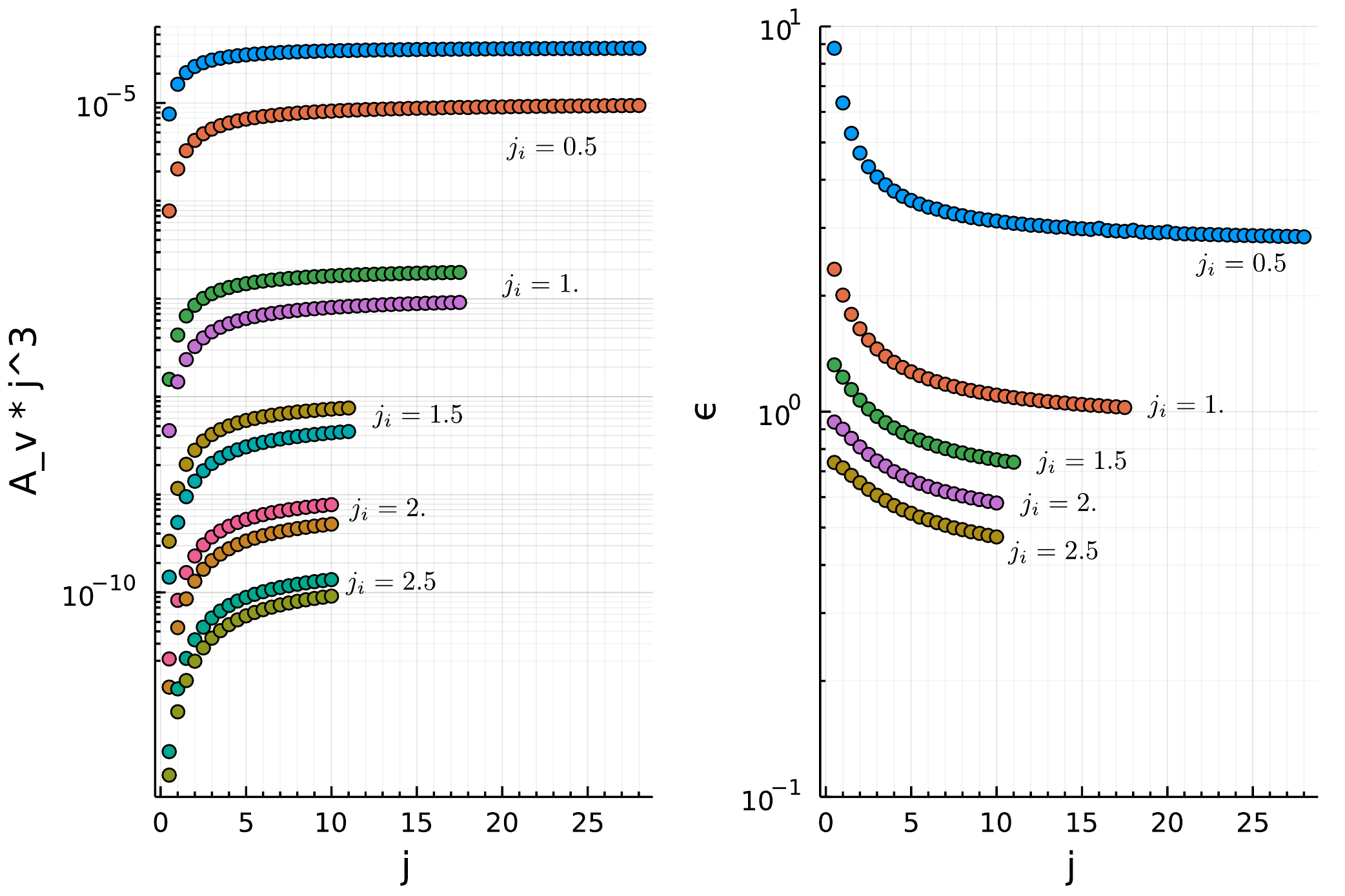}
	\caption{ \label{fig:plot_all_small}
		\textit{Left side:} Semi-classical and full vertex amplitudes $\mathcal{A}_v$ for $j_i = 0.5, \, 1, \, 1.5, \, 2, \, 2.5$, $i \in \{1,\dots,5\}$ and $j_6 = j$ multiplied with $j^3$ with logarithmic scale. The plots are ordered from top to bottom by $j_i = 0.5$, $j_i = 1$, etc. to $j_i = 2.5$, $i \in \{1,\dots,5\}$. The semi-classical amplitude is always larger than the full one. \textit{Right side:} Relative error $\epsilon$ of semi-classical ampltitudes to the full one in logarithmic scale, again $j_i = 0.5$, $j_i = 1$, etc. to $j_i = 2.5$, $i \in \{1,\dots,5\}$ from top to bottom.
	}
\end{figure}

The final set of examples is the most peculiar: we keep five spins $j_1,\dots,j_5$ small and increase $j_6$ as much as numerically possible. For the remaining spins $\{j_i\}_{i=1,\dots,5}$ we choose different values $j=\frac{1}{2},1,\dots,\frac{5}{2}$. The results are summarized in fig. \ref{fig:plot_all_small}. Qualitatively we observe a similar behavior as in the previous examples: full and semi-classical amplitude both show a similar scaling behavior, where the semi-classical amplitude generically overestimates the amplitude for small spins. However, while the gap between full and semi-classical amplitude closes initially as $j_6$ is increased, a finite gap remains (in $\log$-scale) for large $j_6$. As we increase the spins $\{j_i\}_{i=1,\dots,5}$ this gap narrows down further, but does not close for the values of $j_6$ that we have reached. This is reflected also in the relative error, see again fig. \ref{fig:plot_all_small}, which drops from roughly $285\%$ for $j_i=\frac{1}{2}$ to $47\%$ for $j=\frac{5}{2}$. At large $j_6$, increasing $j_6 \rightarrow j_6+1$ typically results in a reduction of the relative error in its third digit. Thus, it appears unlikely that further and further increasing $j_6$ would lead to a regime in which the semi-classical amplitude could well approximate the full one unless we also increase $\{j_i\}_{i=1,\dots,5}$.

To summarize the results for the cuboid amplitude: in agreement with the semi-classical analysis, the full amplitude is strictly positive and shows no oscillatory behavior. Thus, both amplitudes are determined by their scaling behaviour. Under uniform scaling of all spins, we observe a quick convergence between the full and the semi-classical amplitude, but due to high numerical costs we cannot reach large enough spins to prove their equivalence. Additionally, we studied cases in which we keep a subset of spins constant and small, while increasing the remaining ones. Here, we make a few general observations: in general, the fewer spins remain small the better the semi-classical approximation gets in full agreement with the asymptotic expansion. Moreover, the smallness of a few spins can be (at least partially) compensated by making the remaining spins even larger. However, there is also a limit to this, namely if too many spins remain small, increasing the remaining ones cannot compensate this and the semi-classical approximation remains invalid and likely cannot be reached. These findings suggest that the transition between the full quantum regime and the semi-classical one (for a single vertex amplitude) is intricate: it appears plausible that configurations exist for which the semi-classical approximation is valid despite the fact that some spins are small.

\subsection{Frustum results}

The vertex amplitude for frustum intertwiners promises to be more interesting than the cuboid one since we expect an oscillatory behavior from its semi-classical analysis. However, frusta are more restrictive than the cuboid case: 
In addition to depending on just three spins $j_1$, $j_2$ and $k$, these spins cannot be chosen arbitrarily. Firstly, we must ensure that the angle $\phi$, defined by $\cos(\phi) = \frac{j_2 - j_1}{4 k}$, is well-defined. This limits e.g. increasing $j_2$ while keeping $j_1$ and $k$ fixed. 
Additionally, frustum intertwiners do not automatically satisfy the $\text{SU}(2)$ coupling rules: while we can choose $k$ freely, both $j_1$ and $j_2$ have to be either half-integer or integer valued\footnote{The representations at the links of an intertwiner must sum up to an integer. This is automatically satisfied for $k$, since four links carry this spin. The two remaining links carry $j_1$ and $j_2$ respectively.}. Since the numerical costs are similar to the cuboid case, uniform scaling of all spins, which is ideal to showcase the oscillatory behavior of the amplitude, is limited, such that a meaningful comparison and conclusion is difficult. Additionally, we will again explore cases, where some spins remain small and still show a rich oscillatory behaviour.

\begin{figure}
	\includegraphics[width = 0.6\textwidth]{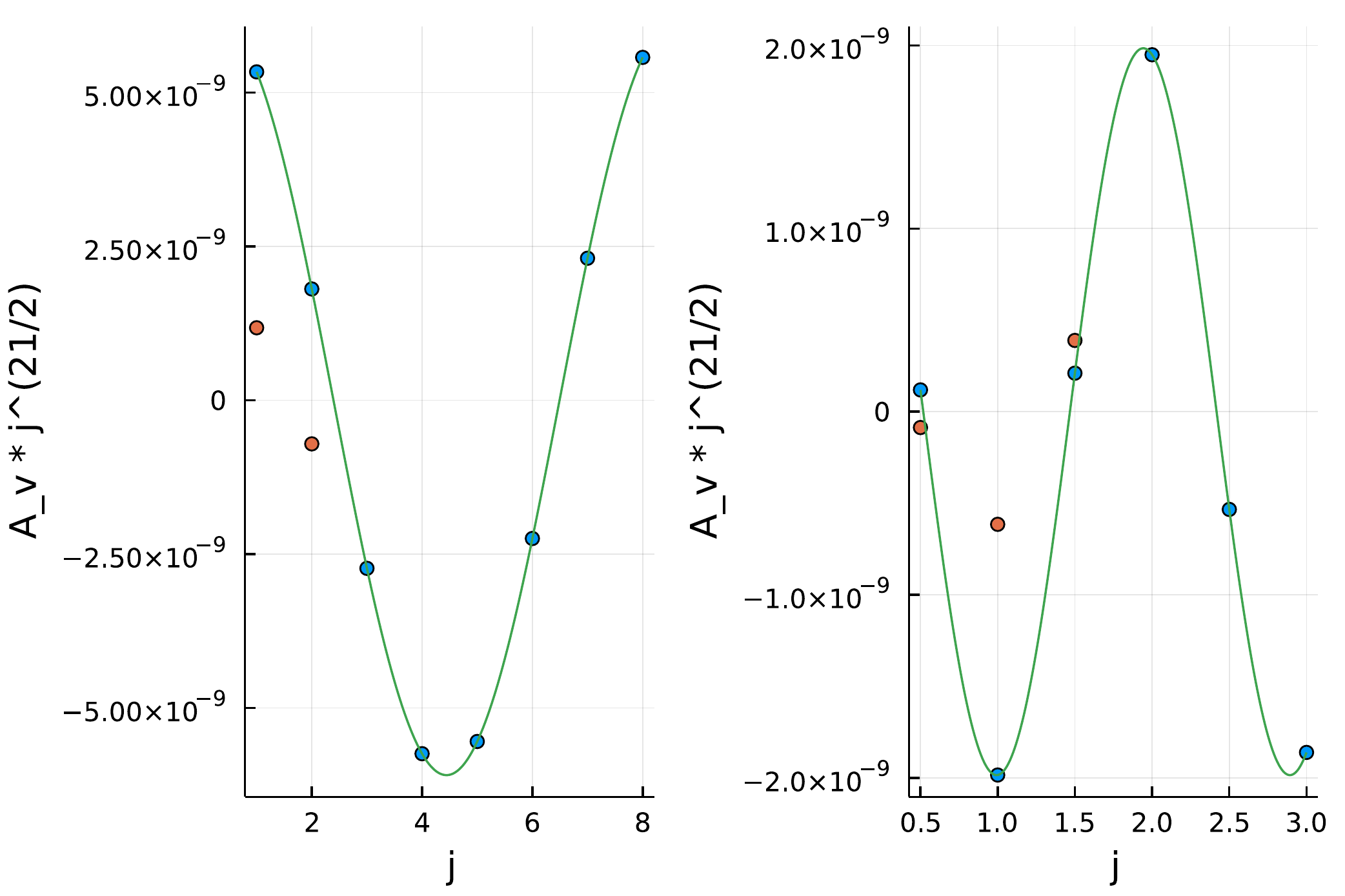}
	\caption{ \label{fig:frusta_uniform_scaling}
		\textit{Left:} Semi-classical (blue) and full vertex amplitude (orange) $\mathcal{A}_v$ multiplied by $j^{\frac{21}{2}}$ of a hyperfrustum with $j_1 = k = j$ and $j_2 = 2j$. Green line shows the semi-classical amplitude for continuous $j$ (ignoring coupling rules).
		\textit{Right:} Semi-classical (blue) and full vertex amplitude (orange) $\mathcal{A}_v$ multiplied by $j^{\frac{21}{2}}$ of a hyperfrustum with $j_1 = k = j$ and $j_2 = 3j$. Green line shows the semi-classical amplitude for continuous $j$ (ignoring coupling rules).
	}
\end{figure}

Let us first discuss uniform scaling of the arguments of the frusta vertex amplitude. Here we do not consider the cases where $j_1 = j_2$, since then $\phi = \frac{\pi}{2}$ and the amplitude reduces to a particular cuboid amplitude\footnote{We note that the results fully agree with the cuboid amplitude studied above.}. To move to a more interesting case, we consider $j_1 = k = j$ and $j_2 = 2j$. This corresponds to a hyperfrustum built from a small cube and a cube twice as large connected by six frusta. The results are presented in fig. \ref{fig:frusta_uniform_scaling}. Unfortunately, due to high numerical costs and constraints from coupling rules, we were not able to compute the vertex amplitude beyond $j=2$\footnote{For $j=3$, we need to compute one cube intertwiner with $j_2 = 6$, which unfortunately proves too costly. The largest intertwiner we were able to compute was for all $j=5.5$, which took more than six days using 8 nodes with 36 cores / 72 threads each.}. Additionally, we study the case $j_1 = k = j$ and $j_2 = 3j$: here the coupling rules are always satisfied such that we can compute more non-vanishing amplitudes (despite larger costs). Still, we can only see half a oscillation of the vertex amplitude.
Due to the lack of data, we cannot draw many conclusions: while we observe a qualitatively similar behavior between the full and the semi-classical amplitude, we also observe differences, namely the scaling behaviour (similar to the cuboid case) and a phase shift in the oscillations. The phase shift is visible in the non-alignment of the roots in both cases. The frequency of oscillations look similar in the case $j_2 = 3j$, but the oscillations are too rapid to be resolved by the spins. Unfortunately we have too little data to conclude a convergence of the amplitudes for large spins.

\begin{figure}
	\includegraphics[width = 0.6\textwidth]{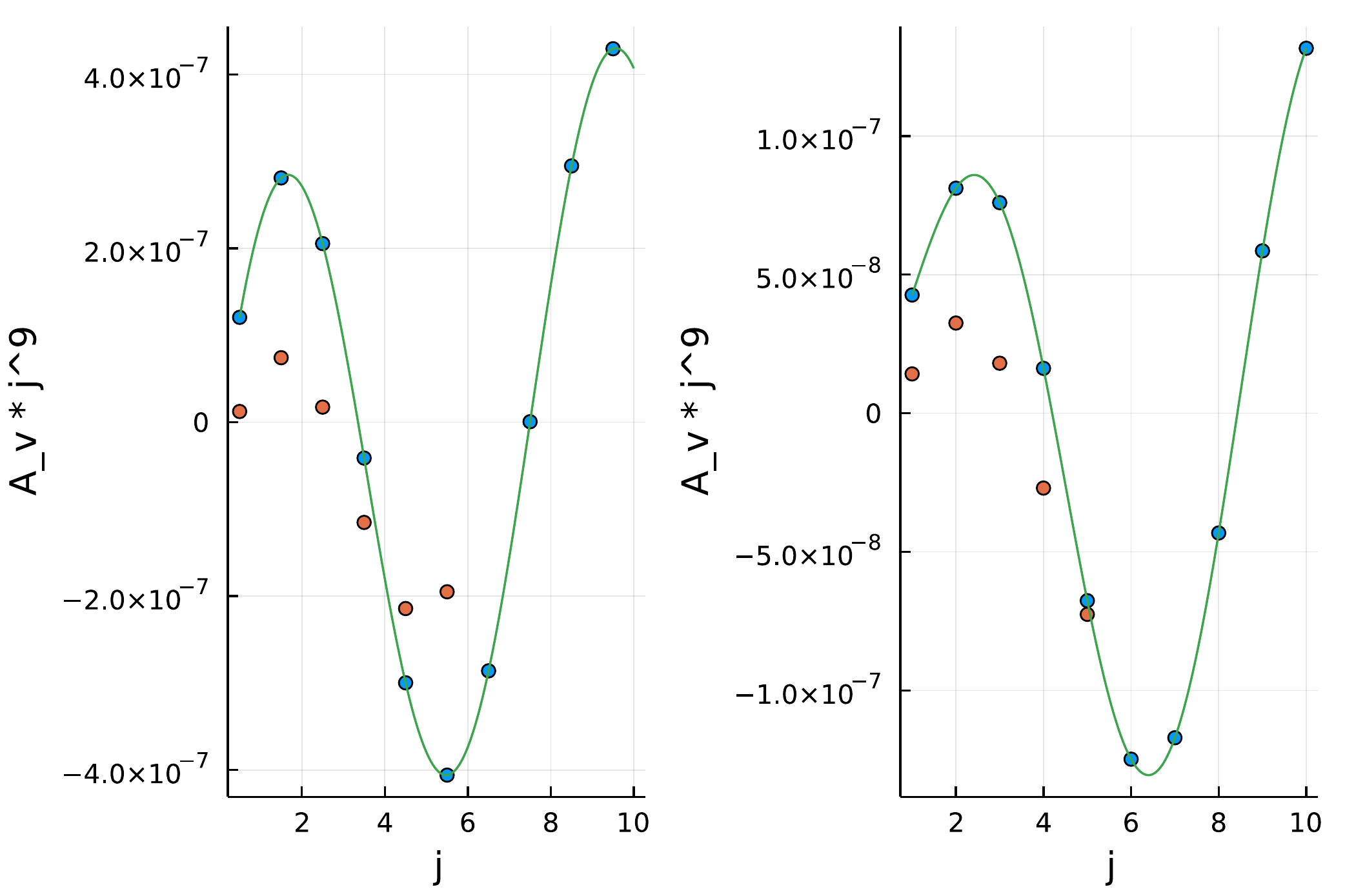}
	\caption{ \label{fig:frusta_small_cube}
		Both sides show the semi-classical (blue) and full vertex amplitude in the frustum case, where we keep $j_1 = i$ small and increase $j_2 = k = j$. The green line shows the semi-classical amplitude for continuous $j$ (ignoring coupling rules). \textit{Left side:} $i = 0.5$. \textit{Right side:} $i = 1$.
	}
\end{figure}

For the next comparison, we keep the spin of the initial cube $j_1$ fixed at a small value and increase $j_2 = k = j$. The results for $j_1=0.5$ and for $j_1=1$ are shown in fig. \ref{fig:frusta_small_cube} left and right side respectively. These cases are interesting, since the semi-classical amplitude shows a clear oscillatory behavior and we can feasibly explore larger values of $j$. Indeed, the full amplitude shows also an oscillatory behavior qualitatively similar to the semi-classical amplitude. The frequency of the full and semi-classical amplitude look similar (as far as one can tell from the data), but we see a phase shift again. Again, as for the previous cases, the absolute value of the full amplitude is smaller than the semi-classical one, yet they approach each other as we increase $j$. Note that since we do not uniformly scale all spins, the frequency of oscillations changes as we increase $j$. Unfortunately, increasing $j$ enough to see convergence is out of reach, but the data look promising. 

In the final set of cases, we consider hyperfrusta where we keep $j_2 - j_1$ fixed and increase $k$, i.e. we keep the size differences of the areas of squares fixed. For large $k$, the frustum intertwiners and the semi-classical amplitude are well-defined, even if $j_1$ and $j_2$ are small. Geometrically, these cases can be understood as follows: if $j_2 - j_1 = 0$, we have the cuboid case again with an angle $\phi = \frac{\pi}{2}$. Increasing this difference (with $4k \geq j_2 - j_1$), decreases $\phi$, which in turn leads a non-vanishing Regge action and an oscillatory behaviour; we observe more rapid oscillations as we increase $j_2 - j_1$. Conversely, increasing $k$ while keeping fixed $j_2 - j_1$ we approach the cuboid case again as $k \rightarrow \infty$, such that we expect oscillations to seize. 

\begin{figure}
	\includegraphics[width = 0.6\textwidth]{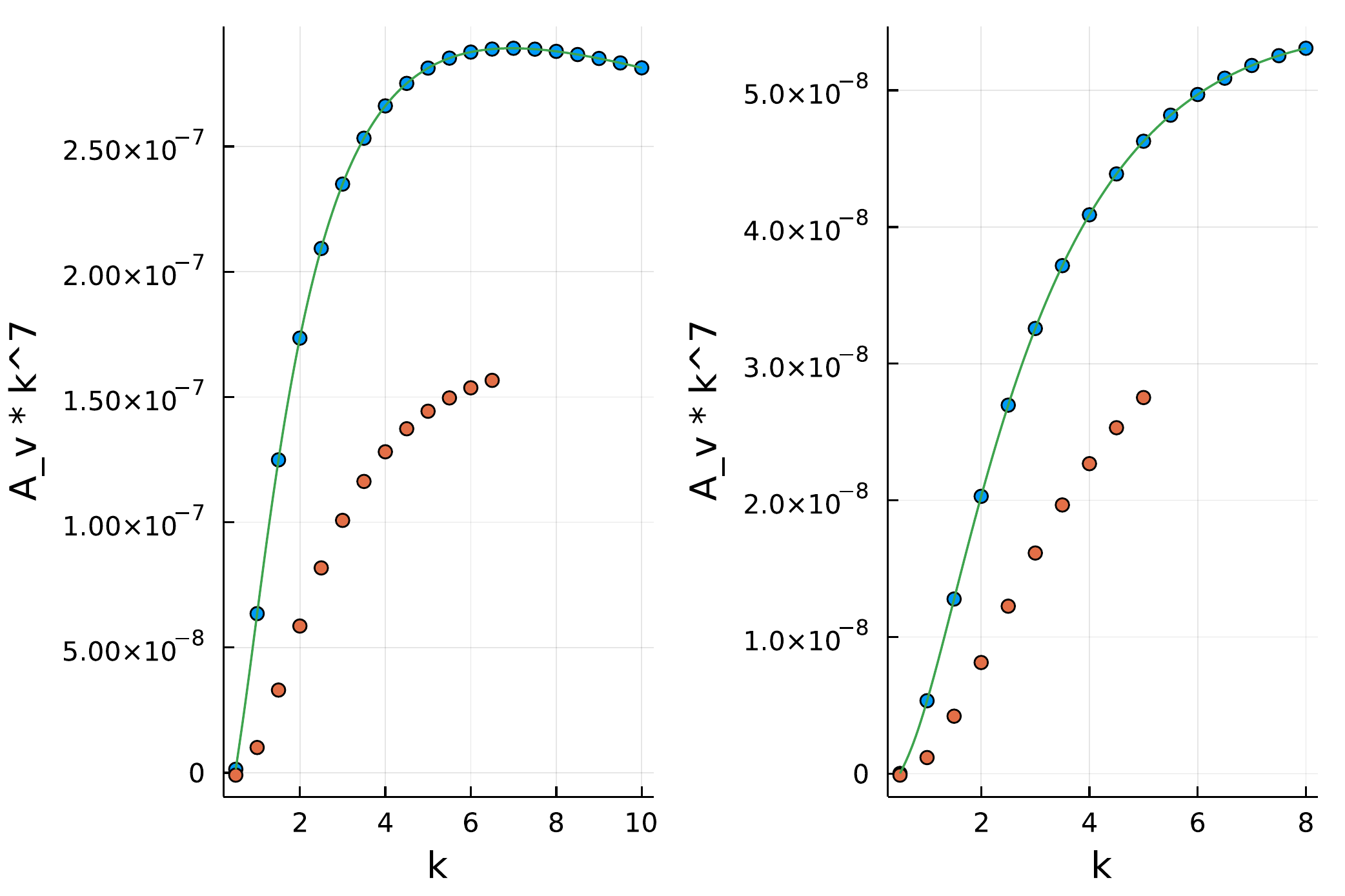}
	\caption{ \label{fig:frusta_diff_1_small}
		Both sides show the semi-classical (blue) and full vertex amplitude in the frustum case, where we increase $k$ and keep $j_1=i$ and $j_2$ small with  $j_2 - j_1 = 1$. The green line shows the semi-classical amplitude for continuous $k$ (ignoring coupling rules). \textit{Left side:} $i = 0.5$. \textit{Right side:} $i = 1$.
	}
\end{figure}

\begin{figure}
	\includegraphics[width = 0.6\textwidth]{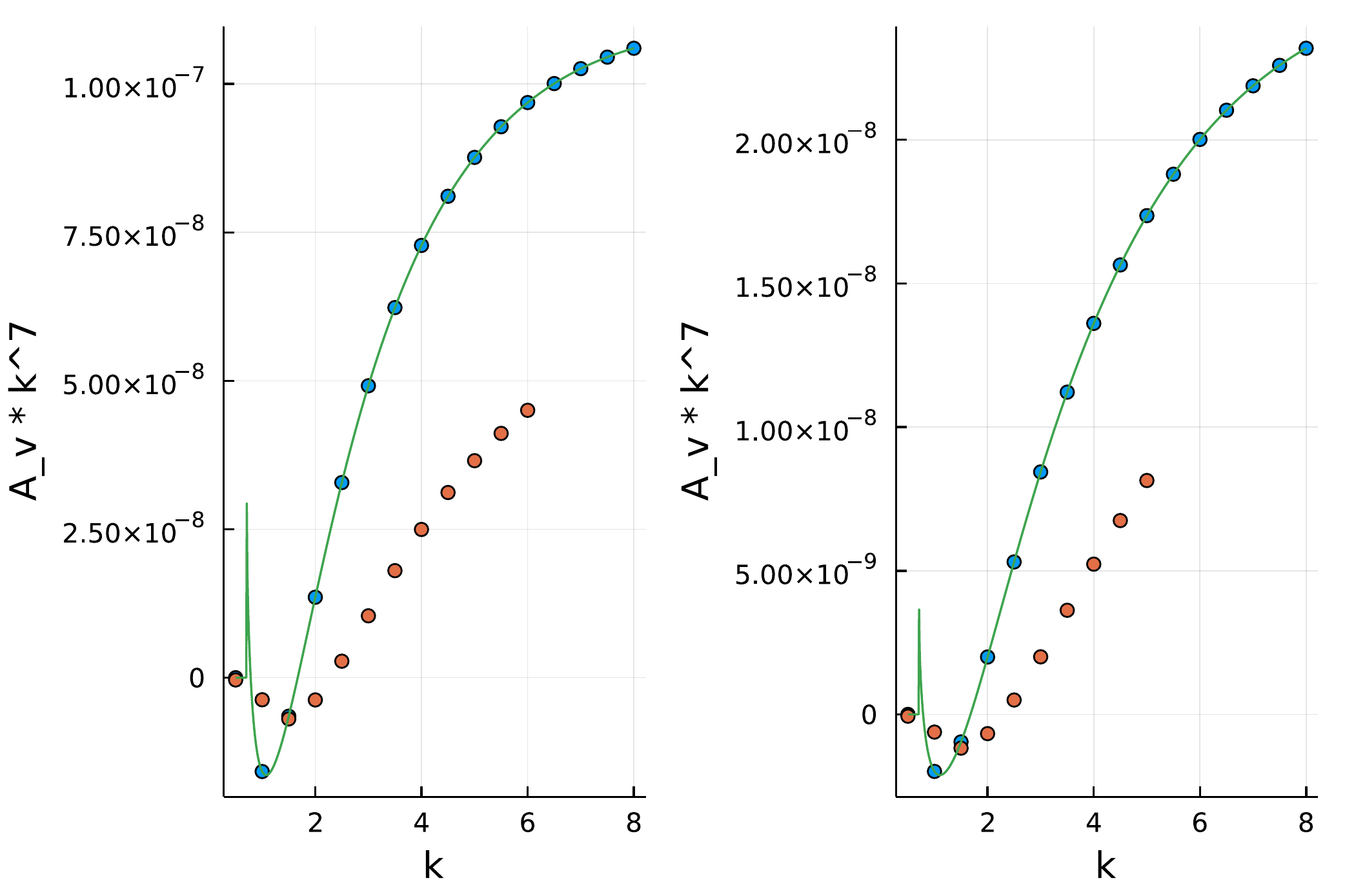}
	\caption{ \label{fig:frusta_diff_2_small}
		Plots of the semi-classical (blue) and full vertex amplitude (orange) in the frustum case for $j_2 - j_1 = 2$, where we keep $j_1 = i$ small and increase $k$. The green line shows the semi-classical amplitude for continuous $k$ (ignoring coupling rules). \textit{Left side:} $i = 0.5$. \textit{Right side:} $i = 1$.
	}
\end{figure}

\begin{figure}
	\includegraphics[width = 0.6\textwidth]{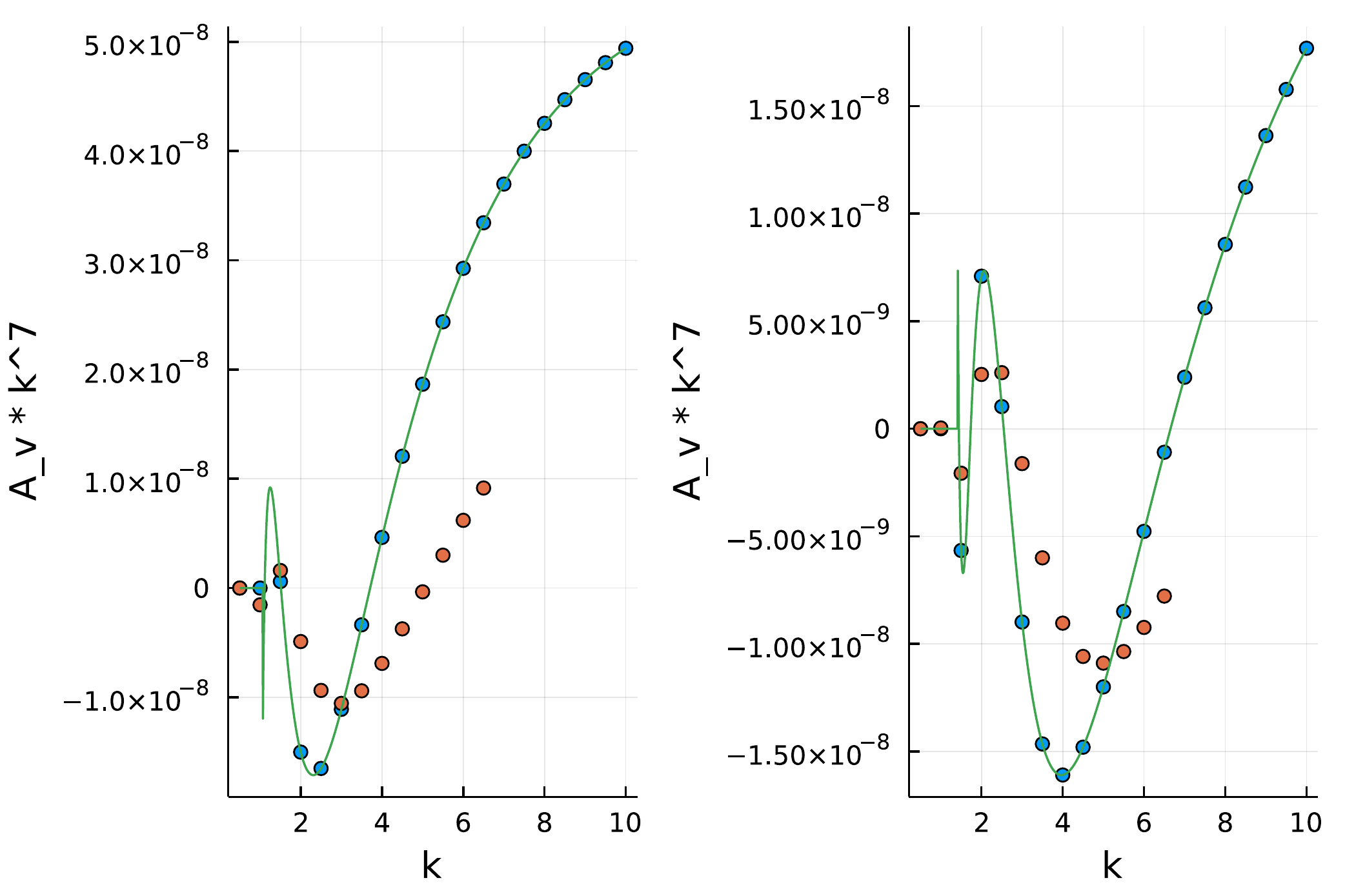}
	\caption{ \label{fig:frusta_diff_3_4}
		\textit{Left:} Semi-classical (blue) and full vertex amplitude (orange) for $j_2 - j_1 = 3$, $j_1 = 0.5$ and we increase $k$. 
		\textit{Left:} Semi-classical (blue) and full vertex amplitude (orange) for $j_2 - j_1 = 4$, $j_1 = 0.5$ and we increase $k$.
		The green line shows the semi-classical amplitude for continuous $k$ (ignoring coupling rules).
		}
\end{figure}

\begin{figure}
	\includegraphics[width = 0.6\textwidth]{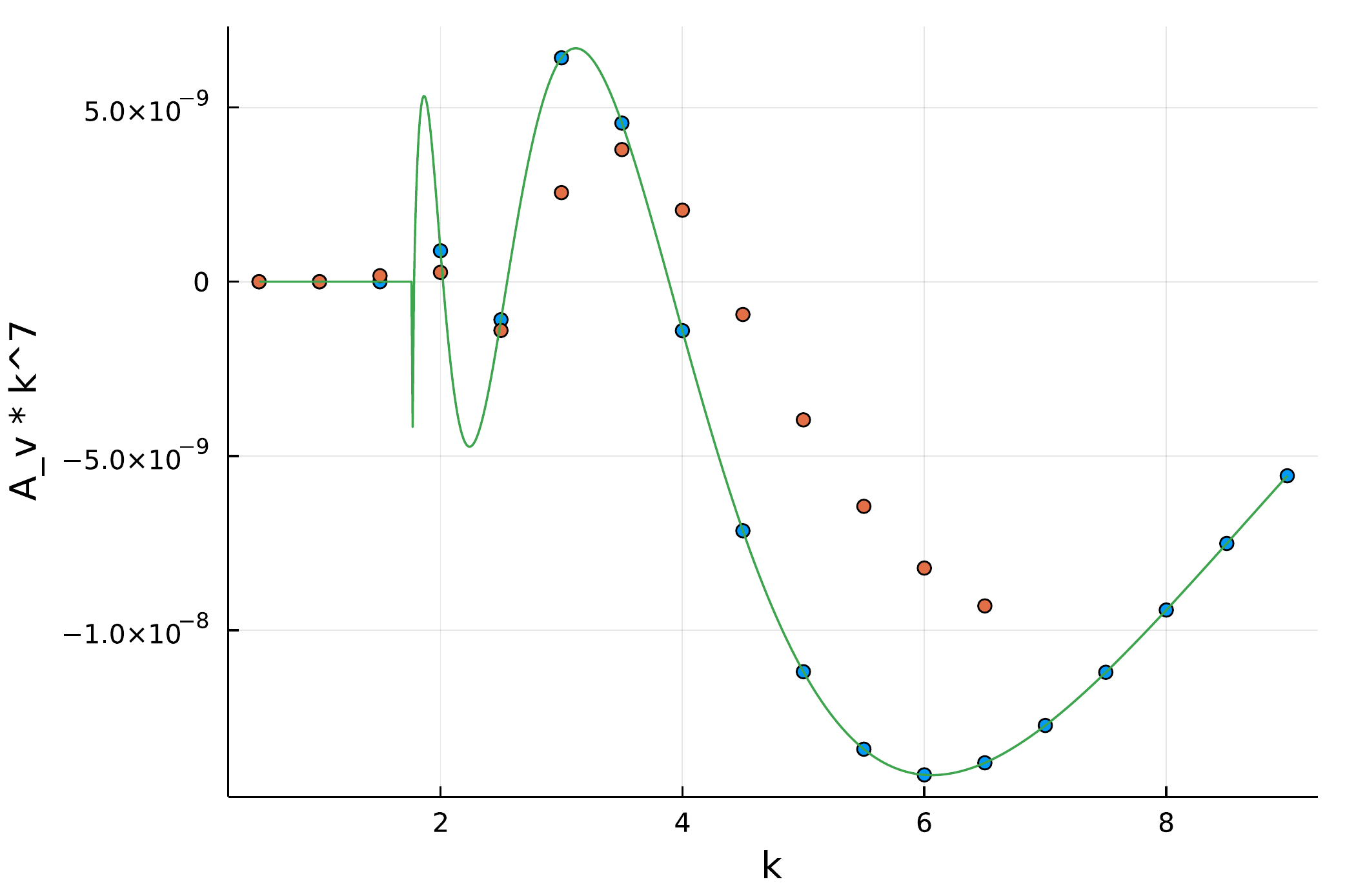}
	\caption{ \label{fig:frusta_diff_5}
		Semi-classical (blue) and full vertex amplitude (orange) for $j_2 - j_1 = 5$, $j_1 = 0.5$ and we increase $k$. The green line shows the semi-classical amplitude for continuous $k$ (ignoring coupling rules).
	}
\end{figure}

Due to the similarity of these cases, we discuss them together. We present the results for $j_2 - j_1 = 1$ in fig. \ref{fig:frusta_diff_1_small} for $j_2 - j_1 = 2$ in fig. \ref{fig:frusta_diff_2_small}, for $j_2 - j_1 = 3$ and $j_2 - j_1 = 4$ in fig. \ref{fig:frusta_diff_3_4} and for $j_2 - j_1 = 5$ in fig. \ref{fig:frusta_diff_5}. Let us first discuss the semi-classical amplitude: at small $k$, we observe more and more rapid oscillations as we increase the difference $j_2 - j_1$. Moreover, we see that the amplitude is about to diverge as $k \rightarrow 0$, but becomes ill-defined and is set to zero. In contrast, as we send $k \rightarrow \infty$ the oscillations seize, the amplitude becomes positive as in the cuboid case.

Qualitatively the full amplitude agrees well with the semi-classical one. At small $k$ it follows the rapid oscillations, i.e. the sign of the amplitude and the sign changes agree, while it simultaneously cures the divergences as we approach $k \rightarrow 0$. This can be nicely seen already at $j_2 - j_1 =2$ in fig. \ref{fig:frusta_diff_2_small}. Furthermore, in cases where the semi-classical amplitude is ill-defined because no critical point exists for this configuration, the full amplitude is well-defined and non-vanishing. Note that since all spins are small, we cannot expect to observe an exponential suppression of the full amplitude due to the non-existence of critical points. For $k \rightarrow \infty$, the frequency of oscillations is slower for the full amplitude, yet in both cases the frequency slows down and eventually seizes, such that both amplitudes approach the cuboid case. As before, the gap for large $k$ between the semi-classical and the full amplitude shrinks if the spins of the frustum are larger, see e.g. fig. \ref{fig:frusta_diff_5}.

In a nutshell, the frustum amplitude is significantly more interesting than the cuboid amplitude, since it tests the oscillatory behaviour of the amplitude in addition to its scaling behaviour. The interpretation of these results are a bit more subtle, however: in general, we observe a good qualitative agreement between both amplitudes. In most cases the frequency of oscillations is close and we only observe a small phase shift. Additionally, the scaling behaviour of the amplitudes is comparable to the cuboid case. Unfortunately, due to the properties of the amplitude, we cannot fully resolve the oscillations to determine the frequency more accurately or we lack the data due to high computational costs. Thus, while we cannot unambiguously prove that the semi-classical amplitude becomes a viable approximation at large spins, our results look promising. Moreover, we also see first indications in the frustum case that the semi-classical approximation might be valid even if some spins remain small (if the other ones are large).



\section{Discussion and Outlook} \label{sec:discussion}

In this article, we present the first numerical calculation of the vertex amplitude of the 4D Euclidean EPRL-FK \cite{Engle:2007wy,Freidel:2007py} in the KKL-extension \cite{Kaminski:2009fm} that allows for higher-valent vertices not dual to a 4-simplex. We compute it for vertices with hypercubic combinatorics and coherent boundary data that correspond to cuboids \cite{Bahr:2015gxa} and frusta \cite{Bahr:2017eyi}. We compare the results to the semi-classical amplitude derived using a stationary phase approximation valid if all spins are large and observe overall good qualitative agreement and convergence of both amplitudes.

If we compare our results to those of the Euclidean 4-simplex in \cite{Dona:2017dvf}, where the equilateral and isosceles 4-simplex are investigated, we observe both similarities and differences. Under uniform scaling, the 4-simplex amplitude shows great agreement of both amplitudes. The frequency of oscillations matches as peaks and roots of the amplitude align well, no phase shift is visible. For large spins the scaling of the amplitudes also agree very well. The expected discrepancy is at small spins, where the full amplitude cures the divergence of its semi-classical counterpart. In our results, we also observe a good convergence of the scaling behaviour, in particular in the cuboid case. From the frusta case, which is the only one with an oscillating amplitude, we see a qualitative agreement, but also slight differences. Since the roots of the amplitudes do not align, one could conclude that there is a phase shift between the amplitudes. However, since we cannot accurately resolve the frequency due to the rapid oscillations of the amplitude, this could also be caused by a difference in frequency (or a combination of both). Due to the high numerical costs for large spins, we cannot say whether both amplitudes agree better as spins increase.

Despite these drawbacks, cuboids and frusta also have an advantage compared to 4-simplices: given a set of 3D normals corresponding to a flat Euclidean 4-simplex together with the ten areas of triangles, one can only uniformly scale the areas. Changing individual spins without adapting the 3D normals leads to exponentially suppressed configurations. In contrast, cuboids and frusta allow us to more freely change single spins, without adapting the boundary data. This allows us to straightforwardly explore new regimes of the vertex amplitude, e.g. where some spins remain small and others become large, to see whether and when the semi-classical amplitude becomes a valid approximation. Additionally, this comes with the benefit of lower computational costs, which allows us to further increase the remaining spins. In particular in the cuboid case, we find several examples where both full and semi-classical amplitude show signs of convergence despite that fact that some spins are small. To compensate for this, the remaining spins need be larger to achieve the same relative error e.g. compared to uniform scaling. We partially see this for frusta as well. Nevertheless, there exists also the opposite situation: if too many spins remain small, the relative error convergences to a non-vanishing value and we cannot compensate for the smallness of spins by increasing the remaining ones. As a final point, these cases give us further insight into the properties of the frusta amplitude: while qualitatively both amplitudes agree well, the full amplitude has a different frequency than the semi-classical one\footnote{Note that under non-uniform scaling the deficit angles of the hyperfrusta change as we increase a subset of spins.}, a feature that is absent for 4-simplices. We do not know the origin of this feature and whether the frequencies agree better if all spins are large. Obtaining these data requires further numerical optimization.


A great obstacle in numerically computing spin foam vertex amplitudes is the rapid growth of computational costs as one increases the representation labels. While the cases for small spins can be performed quickly on modern consumer machines, larger spins require the use of dedicated high performance computing facilities. This fact is already known for spin foam models defined on triangulations \cite{Dona:2017dvf,Dona:2019dkf,Gozzini:2021kbt}, and makes exploring 2-complexes with multiple vertices challenging. In this work, we have observed that the scaling of numerical costs for 2-complexes more general than triangulations is more rapid and severely limits our ability to explore larger spins despite using considerable computational resources. To go further and reach the regime in which the semi-classical amplitude is valid, further optimization is vital, which is however more intricate to implement compared to the triangulation case. In the following we briefly discuss the possibilities.

\begin{itemize}
	\item \textit{Orthonormal spin network basis for intertwiners:}
		The most computationally costly task are the group integrations to compute the coherent 6-valent intertwiners. These group integrations can be avoided by expressing the coherent intertwiners in the spin network basis for a choice of recoupling scheme, which requires three auxiliary $\text{SU}(2)$ spins. To do so one computes the overlap between the spin network basis and the coherent intertwiners; the group integration is then redundant and can be dropped. The coherent vertex amplitude is then given as a contraction of spin network intertwiners times the overlaps of orthonormal and coherent intertwiners.
		
	\item \textit{Express contraction in terms of recoupling symbols:}
		The caveat of the previous point is that it requires us to perform more contractions of intertwiners, namely one contraction for each combination of basis elements of six-valent intertwiners. Together with the fact that this is the second most costly step of the algorithm, this task is daunting and the reason why we did not pursue this direction in this work. Nevertheless, using the spin network basis bears the potential of further optimization as in the 4-simplex case \cite{Dona:2017dvf,Dona:2020tvv}. The $\text{SU}(2)$ $\{15j\}$-symbol can be expressed as a sum over products of five $\{6j\}$-symbols; the computation of $\{6j\}$-symbols is highly optimized and the sum runs over a single spin, which is significantly more efficient than the explicit contraction of intertwiners. For a 2-complex with hybercubic combinatorics such a formula can be derived using $\text{SU}(2)$ recoupling theory, but we expect more internal summations compared to the 4-simplex case.
		
	\item \textit{Sum over intertwiner labels as tensor contractions:}
		The final step to derive the coherent vertex amplitude is to multiply the vertex amplitude in the spin network basis with the overlaps of coherent and spin network intertwiners, and sum over all spin network intertwiner basis elements, here given by three spins each. This operation can be written as the contraction of a tensor network: we interpret the vertex amplitude in the spin network basis as an eight-valent tensor, whose indices label the intertwiner basis elements; in turn, the overlaps are given as vectors. The full amplitude is then the contraction of the vertex amplitude tensor with the eight vectors of overlaps. Such operations can be highly optimized using linear algebra techniques, see e.g. the tensor network community \cite{pfeifer2015ncon}. Such efficient contraction algorithms can e.g. be found in the package \verb|TensorOperations|\footnote{\url{https://github.com/Jutho/TensorOperations.jl}} for \verb|Julia| including GPU support.
\end{itemize}

While these ideas are promising, we cannot estimate how much we could increase the spins compared to the results presented in this article. However, all the above-mentioned optimizations are or can be implemented in the algorithm for 4-simplices as well, which will still be less costly computationally compared to more general 2-complexes. Thus, spin foam models for 4-simplices promise the best chances to be feasibly numerically explored, such that we can identify the regime in which the semi-classical approximation is valid. Additionally there are further arguments in favor of 4-simplices: in \cite{Bahr:2017ajs}, it was explicitly shown that the EPRL-FK model defined on 2-complexes more general than triangulations does not impose the so-called volume simplicity constraint. In a 4-simplex, this constraint requires that the 4D volume spanned by two bivectors, which are assigned to triangles that only share one vertex of the 4-simplex, is the same for any choice of such bivectors. Indeed, this constraint is automatically implemented classically once the other constraints (diagonal and cross-simplicity \cite{Engle:2007qf}) are enforced and thus it is not explicitly implemented in the EPRL-FK model. These works suggest that the EPRL-FK models defined on triangulations and non-triangulations are not the same, where the former appears as a more suitable candidate for a theory of quantum gravity. Moreover, it is not clear how volume simplicity could be suitably implemented in spin foams defined on general 2-complexes \cite{Assanioussi:2020fml}.

Nevertheless, we can still draw several interesting conclusions from this work. The original motivation to define the restricted models was to explore a subset of the gravitational path integral of spin foam models to study its renormalization and observables. Indeed, first indications for a UV-attractive fixed point were found in \cite{Bahr:2016hwc,Bahr:2017klw,Bahr:2018gwf} as well as a study of the spectral dimension of the cuboids \cite{Steinhaus:2018aav}. Both examples are highly sensitive to the scaling behavior of the spin foam amplitudes (including face and edge amplitudes), such that the modified scaling behavior at small spins could modify these results. For example, the spectral dimension of cuboid spin foams is solely determined by the scaling behavior of the amplitude, which we thus expect to change at small spins. We hope that similar calculations become possible soon for the full model defined on triangulations thanks to the development of powerful numerical tools \cite{Dona:2017dvf,Dona:2019dkf,Gozzini:2021kbt} as well as using effective spin foam models \cite{Asante:2020iwm,Asante:2020qpa,Asante:2021zzh} and Lefshetz-thimble Monte Carlo techniques \cite{Han:2020npv}.

\section*{Acknowledgements}
C.A. would like to thank Erik Schnetter for discussions on implementing parallelization in \verb|Julia|.
S.St. would like to thank Benjamin Bahr and Sebastian Kl\"oser for early discussions about how to compute the cuboid vertex amplitude.

S.St. is funded by the Deutsche Forschungsgemeinschaft (DFG, German Research Foundation) - Projektnummer / project-number 422809950. This research was in part supported by Perimeter Institute for Theoretical Physics. Research at Perimeter Institute is supported in part by the Government of Canada through the
Department of Innovation, Science and Economic Development Canada and by the Province of Ontario through the
Ministry of Colleges and Universities.

\appendix

\section{Notation for intertwiners and their contractions} \label{app:notation}

\begin{figure}
	\includegraphics[width = 0.4\textwidth]{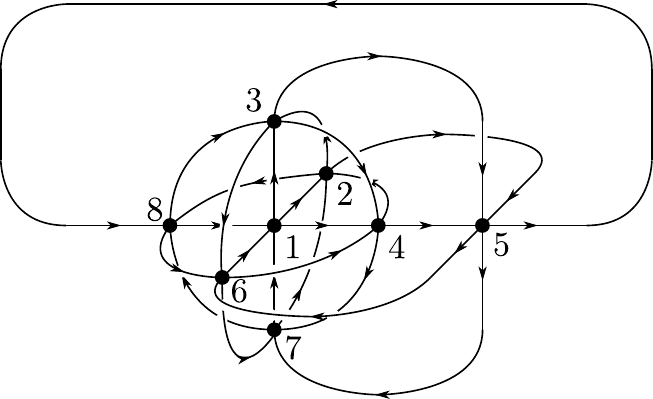} \quad \quad \quad \quad
	\includegraphics[width = 0.225\textwidth]{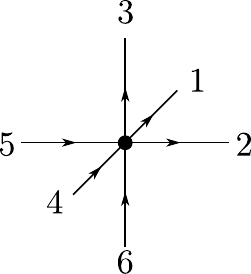}
	\caption{ \label{fig:app_snw}
		\textit{Left:} Enumeration of intertwiners in the vertex amplitude. \textit{Right:} Enumeration of indices of all intertwiners.
	}
\end{figure}

In the code we are using the following notation for the intertwiners and the vertex amplitude: firstly, we enumerate the intertwiners as shown on the left of fig. \ref{fig:app_snw}. Secondly, we enumerate the indices of each intertwiner as on the right of fig. \ref{fig:app_snw}. Following this notation, the first index of intertwiner $1$ gets contracted with the fourth index of intertwiner $2$.

The final vertex amplitude is then given by:
\begin{align} \label{eq:vertex_amplitude_code}
	\mathcal{A}_v = & \sum_{\{m_i\}} \iota_1(m_1,m_2,m_3,m_4,m_5,m_6) \; \iota_2(m_7,m_8,m_9,m_1,m_{10},m_{11}) \nonumber \\
	& \times \iota_3(m_9,m_{12},m_{13},m_{14},m_{15},m_3) \; \iota_4(m_8,m_{16},m_{12},m_{17},m_2,m_{18}) \nonumber \\
	& \times \iota_5(m_7,m_{19},m_{13},m_{20},m_{16},m_{21}) \; \iota_6(m_4,m_{17},m_{14},m_{20},m_{22},m_{23}) \nonumber \\
	& \times \iota_7(m_{11},m_{18},m_6,m_{23},m_{24},m_{21}) \; \iota_8(m_{10},m_5,m_{15},m_{22},m_{19},m_{24}) \quad .
\end{align}
Here we have written the magnetic indices of the intertwiner as arguments to make them more readable. In total the sum runs over 24 indices, each within the bounds given by the respective $\text{SU}(2)$ representation.

We optimize this summation by using the properties of intertwiners: for seven intertwiners we can fix one magnetic index as a function of the others if this solution is within the allowed bounds. Concretely we fix $m_1$, $m_7$, $m_{12}$, $m_{16}$, $m_{20}$, $m_{22}$ and $m_{24}$.

The calculation is written as a \verb|for|-loop over all non-fixed magnetic indices. We parallelize this operation by splitting five variables, $m_2$ to $m_6$, off into an outer loop, which we parallelize. The results are added up using \verb|atomic| addition to synchronize the contributions from different threads.

\bibliography{references.bib}

\end{document}